\documentclass[aps,pra,
amsmath,amssymb,
twocolumn,
groupedaddress,
longbibliography,
floatfix]{revtex4-2}

\usepackage[utf8]{inputenc}
\usepackage[T1]{fontenc}
\usepackage{booktabs,array}
\usepackage{amsthm}
\newtheorem{proposition}{Proposition}
\newtheorem{theorem}{Theorem}
\usepackage{hyperref}
\usepackage{graphicx}
\usepackage{xcolor}
\makeatletter

\let\inputrows\@@input
\makeatother
\usepackage{algorithm}
\usepackage{algpseudocode}

\makeatletter
\def\ftype@algorithm{8}
\providecommand*\toclevel@algorithm{0}
\makeatother
\hypersetup{hidelinks,hypertexnames=false}

\algrenewcommand{\algorithmicrequire}{\textbf{Input:}}
\algrenewcommand{\algorithmicensure}{\textbf{Output:}}

\begin{document}
\title{
From optimization to construction: Expanding quantum measurement designs
}

\author{Taotao Zhao}
\affiliation{School of Electronic Science and Engineering, Southeast University, Nanjing, 211189, China}
\author{Qianjun Yao}
\affiliation{School of Electronic Science and Engineering, Southeast University, Nanjing, 211189, China}

\author{Mengdi Sun}
\altaffiliation[Also at ]{Jiangsu Key Laboratory for Undersea Communications and Sensing, Changshu, 215500, China}
\affiliation{School of Electronic Science and Engineering, Southeast University, Nanjing, 211189, China}

\author{He Li}
\email{Contact author: helix@seu.edu.cn}
\affiliation{School of Electronic Science and Engineering, Southeast University, Nanjing, 211189, China}

\begin{abstract}
Estimating molecular energies on quantum computers requires repeated measurements of many Pauli terms, making sampling a major resource cost.
Measurement design chooses which compatible terms are measured together and allocates a finite shot budget among bases to minimize the variance of an energy estimator.
We introduce variance-guided measurement group construction (VGMC), which reduces variance beyond optimization on existing measurement bases by changing which Hamiltonian terms share measurement shots.
After including every Hamiltonian term measurable in each existing basis, we jointly optimize estimator coefficients and shot allocations to separate inefficient use of these measurements from limitations of their measurement combinations.
we compare a candidate's benefit with the cost of withdrawing shots from existing bases, and use the resulting scores to guide greedy construction of qubit-wise commuting groups followed by joint re-optimization.
With exact covariances, 
VGMC outperformed the state-of-the-art measurement design methods across the molecular benchmarks \textcolor{black}{where it introduced additional measurement bases}, achieving variance reductions of 18.19\%–23.38\%.
The gains persisted with approximate covariances, and simulations with
finite shots indicated lower estimated shot requirements near chemical accuracy. 
These results show that optimizing how measurements are combined can also guide which terms are measured together, offering a route to reducing the sampling overhead of variational quantum simulation while  retaining local measurements on current and near-term devices.  
\end{abstract}
\maketitle

\section{Introduction}\label{sec-introduction}

In the variational quantum eigensolver (VQE) for molecular electronic structure~\cite{cao2019chemistry,mcardle2020chemistry}, repeated energy estimates guide classical optimization of a trial state's parameters~\cite{peruzzo2014vqe,mcclean2016theory,cerezo2021variational,bharti2022noisy}.
A fermion-to-qubit mapping expresses the Hamiltonian as a weighted sum of Pauli operators \cite{bravyi2002fermionic,seeley2012bravyi}, whose expectation values are estimated from repeated cycles of state preparation and measurement, or shots.
The total time spent on state preparation, basis rotations and readout across the many shots required for a precise energy estimate can make energy measurement a practical bottleneck~\cite{wecker2015solving,gonthier2022roadblock,tilly2022vqe}.
Reducing the variance of an  energy estimator improves
precision at a fixed shot budget or reduces the shots needed for a prescribed statistical precision \cite{crawford2021efficient}.
The measurement design problem is therefore to choose measurement bases and allocate a finite shot budget among them so as to minimize the variance of an energy estimator.

Grouping compatible Pauli terms for simultaneous measurement was an early focus of measurement design for VQE~\cite{verteletskyi2020minimumclique}.
A commonly used criterion for such compatibility is qubit-wise commutativity (QWC), which permits simultaneous measurement using only local basis rotations but restricts which terms can be grouped together.
Relaxing this criterion to
fully commuting (FC) allows more terms to share a measurement basis and can reduce the number of bases, but generally requires entangling basis changes that increase circuit depth and exposure to gate errors~\cite{yen2020measuring,hamamura2020efficient}.
Methods that account for gate errors and connectivity \cite{burns2025galic}, limit diagonalization depth \cite{dalfavero2025commutativity}, or adapt diagonalization circuits to hardware connectivity \cite{miller2024hardware} address this tradeoff.
We therefore restrict the measurement design to QWC groups to retain local basis rotations.
Within this class, randomized or derandomized designs select Pauli bases directly~\cite{huang2020classical,huang2021derandomization,hadfield2022biased,elben2023randomized}, while grouping methods organize compatible Hamiltonian terms.
Graph coloring seeks small group counts, and sorted insertion (SI) builds compatible groups in decreasing order of Hamiltonian coefficient magnitude.
The selected 
groups also affect 
subsequent estimator optimization.
SI incorporating covariance information produced initial disjoint groups that yielded lower estimator variances after iterative coefficient splitting (ICS) in molecular tests with FC measurements~\cite{huidobro2026initializations}.

Overlapped grouping allows the same Pauli operator to be estimated from several measurement bases \cite{wu2023ogm}, and an overlapped extension of SI inserts previously assigned terms into other compatible groups.
Each shared term then has several sample means available for estimating its expectation. Its fixed Hamiltonian coefficient can be split among those sample means, provided that the assigned coefficients sum to the original coefficient. These estimator coefficients determine how the measurements are combined while preserving an unbiased energy estimate.

Each group's contribution to the estimator variance is  the variance of its weighted sum of Pauli measurement results from a single shot, divided by its shot count. The estimator coefficients determine the variance of each weighted sum through both individual variances and covariances among the measured terms. Changing the coefficient split therefore changes the best shot allocation, while the best split depends on the precision supplied by each group's shot count. Joint optimization minimizes variance over these coupled choices at a fixed total budget. For fixed coefficients, optimal shot allocation is proportional to the standard deviations of these weighted sums.
Coefficient splitting and ICS use this freedom to improve the estimator \cite{yen2023deterministic}.
Post-hoc repacking makes further use of existing measurement results by including Hamiltonian terms measurable in the selected bases but omitted from the original groups \cite{rowland2026overlapped}. Ad-hoc repacking prioritizes insertions using coefficient magnitudes and counts of the groups containing each term, with its result on strict variance reduction established under a model that assumes zero covariances \cite{rowland2026overlapped}.

Completing each selected basis with all measurable Hamiltonian terms and jointly optimizing coefficients and shots exhausts the estimation freedom on those bases.
Additional bases can make other combinations measurable together, creating new options for coefficient assignment.
Their value depends on the whole estimator because, at a fixed budget, shots assigned to a new basis are withdrawn from the existing ones.
We therefore ask whether the optimized estimator can determine when transferring a small fraction of shots to a candidate basis would reduce variance. 
If so, this information could guide the construction of additional bases by identifying compatible term combinations that make more effective use of the shot budget.

To address this question, we develop variance-guided measurement group construction (VGMC) to connect estimator optimization with measurement group construction. 
Within multiresponse $c$-optimal design, we eliminate estimator coefficients and optimize the shot fractions for the current groups.
For a fixed positive-definite covariance model, the optimality conditions yield a candidate score. 
Comparing this score with the current objective value determines whether transferring shots from existing groups to the candidate can improve the current design.
The score increments guide greedy QWC group construction, and joint reoptimization updates the estimator used in subsequent searches.
Continuous optimization thus supplies information for choosing additional measurement combinations.

The molecular comparisons show variance reductions beyond the jointly optimized initial designs.
Comparisons with random construction at matched addition counts support the value of selection guided by the estimator.
The gains also persisted with approximate covariances from configuration interaction with singles and doubles (CISD) and with finite, integer shot allocations.
These findings establish that an optimized design can supply guidance for improving the measurement combinations themselves, offering a route to reducing the sampling overhead of variational quantum algorithms while retaining local measurements.

Section~\ref{sec-model} defines the estimation problem, and Section~\ref{sec-reduction} develops coefficient assignment, shot allocation, and candidate scoring.
Section~\ref{sec-construction} presents group construction and its stopping conditions.
Section~\ref{sec-experiments} evaluates molecular performance, matched count controls, CISD-based design, and finite shot energy estimation.
Section~\ref{sec-conclusion} discusses the implications and remaining limitations.

\section{Problem formulation}\label{sec-model}

\subsection{Measurement groups and estimators}

We consider energy estimation at a fixed step of VQE. Independent preparations of a state $\rho$ of $N_q$ qubits are used
to estimate $E_{\rho} = \operatorname{Tr}(\rho H)$ for the Hamiltonian
\begin{equation}
H = c_{0}I + \sum_{j = 1}^{N_{P}}c_{j}P_{j}.\label{eq-hamiltonian}
\end{equation}
After repeated operators have been combined,
the $P_{j}$ are distinct nonidentity Pauli operators with known nonzero real coefficients $c_{j}$, 
while the real identity coefficient $c_{0}$ requires no measurement.
Here $N_P$ is the number of retained terms, $H,\rho,P_j\in\mathbb C^{2^{N_q}\times2^{N_q}}$, and $I=I_{2^{N_q}}$ is the identity on the qubit Hilbert space.
Throughout, $I_d$ denotes the $d\times d$ identity matrix and vectors are column vectors.
The state is held fixed throughout the measurement design problem.

Although a local Pauli basis specifies an X, Y, or Z measurement on each qubit, current quantum processors typically provide native readout only in the computational (Z) basis, with X and Y measurements implemented by local basis rotations prior to readout.
Each output bit is mapped to the eigenvalue of the selected Pauli observable, with 0 corresponding to $+1$ and 1 to $-1$, so one shot yields a vector of measured single-qubit Pauli eigenvalues in $\{\pm1\}^{N_q}$ \cite{huang2021derandomization}.
The measurement outcome of a compatible Pauli operator is the product of these eigenvalues over its nonidentity factors.
For each selected basis, we include every compatible Hamiltonian term in its group. 
The basis defines the measurement, and the group records the terms available for coefficient assignment.
Let $\mathcal{G} = \left\{ G_{1},\ldots,G_{L} \right\}$ denote the $L$ groups, with $G_\alpha\subseteq\{1,\ldots,N_P\}$ for $\alpha=1,\ldots,L$ and every term covered at least once.

In a disjoint grouping, each term contributes through one measurement batch with coefficient $c_j$. 
Overlap allows its expectation to be estimated from several batches.
Following 
coefficient splitting~\cite{yen2023deterministic}, we assign a real coefficient $x_{\alpha j}$ to operator $P_j$ in each group $G_\alpha$ with $j\in G_\alpha$.
For the sample mean ${\hat{p}}_{\alpha j}$ obtained from $m_{\alpha} > 0$ shots, the estimator and coefficient condition are
\begin{equation}
    \hat{E} = c_{0} + \sum_{\alpha:m_{\alpha} > 0}\sum_{j \in G_{\alpha}}x_{\alpha j}{\hat{p}}_{\alpha j},
    \sum_{\alpha:j \in G_{\alpha}}x_{\alpha j} = c_{j}.\label{eq-estimator}
\end{equation}
The coefficient condition preserves the Hamiltonian expectation for arbitrary Pauli expectations. A group receiving no shots has zero assigned coefficients.
We optimize over this family of linear unbiased estimators formed from Hamiltonian terms.

Write $c=(c_1,\ldots,c_{N_P})^T\in\mathbb R^{N_P}$ and let $d_\alpha=|G_\alpha|$.
Collect each group's coefficients into $x_\alpha\in\mathbb R^{d_\alpha}$, using a fixed ordering within each group for all its vectors and matrices.
The stacked vector $x=(x_1^T,\ldots,x_L^T)^T\in\mathbb R^{N_x}$ has $N_x=\sum_\alpha d_\alpha$ entries.
We use $S_{\alpha} \in \left\{ 0,1 \right\}^{d_\alpha \times N_{P}}$ to select the entries indexed by $G_{\alpha}$ from a vector whose components follow the ordering $P_{1},\ldots,P_{N_{P}}$.
The coefficient condition then becomes
\begin{equation}
\sum_{\alpha}S_{\alpha}^{T}x_{\alpha} = c.\label{eq-unbiasedness}
\end{equation}
The transpose places the local coefficients in an $N_P$-dimensional vector whose entries follow the order $P_1,\ldots,P_{N_P}$, so Eq.~\eqref{eq-unbiasedness} enforces the coefficient condition for every term.

\subsection{Estimator variance and optimization} \label{sec-estimator-variance}
\textcolor{black}{
Within group $G_\alpha$, the covariance matrix $\Gamma_\alpha^{\mathrm{true}}\in\mathbb R^{d_\alpha\times d_\alpha}$ has entries $\left( \Gamma_{\alpha}^{\text{true}} \right)_{ij} = \operatorname{Tr}(\rho P_{i}P_{j}) - \operatorname{Tr}(\rho P_{i})\operatorname{Tr}(\rho P_{j})$ for $i,j \in G_{\alpha}$, in the group's fixed ordering.}
The variance per shot of the weighted outcome is $x_{\alpha}^{T}\Gamma_{\alpha}^{\text{true }}x_{\alpha}$, and its sample mean has this variance divided by $m_\alpha$.
Batches measured in different bases are independent even when their groups share an operator, because they use different state preparations.

For a total of $M$ shots with equal cost per shot, write $a_{\alpha} = m_{\alpha} / M$  and collect these fractions into $a=(a_1,\ldots,a_L)^T\in\mathbb R^L$.
The variance of the energy estimator is
\begin{equation}
\operatorname{Var}(\hat{E}) = \frac{1}{M}\sum_{\alpha:a_{\alpha} > 0}\frac{x_{\alpha}^{T}\Gamma_{\alpha}^{\text{true }}x_{\alpha}}{a_{\alpha}}.\label{eq-variance}
\end{equation}
Coefficient assignment and shot allocation are therefore coupled.
The coefficients determine each group's variance, while its shot count determines the precision with which those coefficients contribute to the energy estimate.
Covariances enter through products of the assigned coefficients, so their effect depends on both signs and magnitudes.

For measurement design, use fixed positive-definite matrices $\Gamma_\alpha\in\mathbb R^{d_\alpha\times d_\alpha}$, obtained from exact or approximate covariance information.
We write $\Gamma_\alpha\succ0$ for positive definiteness and $\Gamma_\alpha\succcurlyeq0$ for positive semi-definiteness.
We relax the shot fractions to the closed simplex $\Delta_L=\{a\in\mathbb R^L:a_\alpha\ge0,\ \sum_{\alpha=1}^L a_\alpha=1\}$.
When $a_\alpha=0$, the contribution of group $G_\alpha$ to the objective is zero if $x_\alpha=0$ and infinite otherwise, as formalized in Appendix~\ref{app-coefficient-assign}.
The joint optimization problem is
\begin{equation}
\begin{aligned}
    \Phi_{\mathcal{G}}^{\star} = \min\limits_{a \in \Delta_{L},\sum_{\alpha}S_{\alpha}^{T}x_{\alpha} = c}\Psi(x,a),\Psi(x,a) = \sum_{\alpha:a_{\alpha} > 0}\frac{x_{\alpha}^{T}\Gamma_{\alpha}x_{\alpha}}{a_{\alpha}}.\label{eq-joint}
\end{aligned}
\end{equation}
The model predicts estimator variance $\Psi(x,a)/M$. When its covariances match those of the measured state, $\Psi(x,a)=M\operatorname{Var}(\hat E)$.
Here $\Phi_{\mathcal G}^\star$ is the scalar minimum of $\Psi$ over both coefficient assignments and shot fractions for the fixed group set $\mathcal G$. We write $(x^\star,a^\star)$ for a minimizing pair.
Holding the covariance model fixed while enlarging the group set isolates the improvement available from additional measurement choices after optimizing all coefficients and shot fractions.

\section{Optimization and group evaluation}\label{sec-reduction}

The joint problem in Eq.~\eqref{eq-joint} of Section~\ref{sec-estimator-variance} couples the estimator coefficients $x$ and shot fractions $a$ for a fixed group set and covariance model.
It is a special case of multiresponse $c$-optimal design, which chooses measurements to minimize the variance of a specified linear combination of quantities estimated jointly~\cite{sagnol2011multiresponse}.
Here the linear combination is the energy expectation, with coefficients $c$.
We derive two equivalent formulations by eliminating one set of variables at a time.
Eliminating the shot fractions gives a second-order cone program (SOCP) in the coefficients.
Eliminating the coefficients gives an optimization over shot fractions in which every allocation uses its optimal estimator.
The derivative of the reduced allocation objective provides the scores used to evaluate additional measurement groups.

\subsection{SOCP}\label{sec-socp}

A second-order cone program minimizes a linear objective subject to linear equalities and constraints that bound Euclidean norms of affine vector expressions by scalar affine expressions~\cite{lobo1998applications}.
An affine expression is a linear expression plus a constant.
For a positive integer $d$, a scalar $\tau\in\mathbb R$, and a column vector $z\in\mathbb R^d$, the standard second-order cone is
\begin{equation}
\mathcal Q_{d+1}
=\{(\tau,z)\in\mathbb R\times\mathbb R^d:\|z\|_2\le \tau\}.
\label{eq-soc-definition}
\end{equation}
Thus a constraint of the form $(\tau,z)\in\mathcal Q_{d+1}$ bounds the Euclidean norm of a vector by a scalar and implies $\tau\ge0$.
Here $\|z\|_2=(z^Tz)^{1/2}$.
To put Eq.~\eqref{eq-joint} into this form, we first optimize the shot fractions for each feasible coefficient assignment.

Let $\Gamma_\alpha^{1/2}\in\mathbb R^{d_\alpha\times d_\alpha}$ be the symmetric positive-definite square root of $\Gamma_\alpha$.
This matrix satisfies $(\Gamma_\alpha^{1/2})^T\Gamma_\alpha^{1/2}=\Gamma_\alpha$.
For coefficients satisfying $\sum_\alpha S_\alpha^Tx_\alpha=c$, define
\begin{equation}
r_\alpha(x)=\sqrt{x_\alpha^T\Gamma_\alpha x_\alpha}
=\|\Gamma_\alpha^{1/2}x_\alpha\|_2,
\ F(x)=\sum_\alpha r_\alpha(x).
\label{eq-group-standard-deviation}
\end{equation}
Here $r_\alpha(x)$ is the standard deviation predicted by the covariance model for the weighted sum of Pauli measurement outcomes from one shot in group $G_\alpha$, with weights $x_{\alpha j} $.
Since $c\ne0$, at least one group coefficient vector is nonzero, and positive definiteness ensures $F(x)>0$.
For any allocation compatible with these coefficients, the Cauchy--Schwarz inequality and $\sum_\alpha a_\alpha=1$ give
\textcolor{black}{
\begin{equation}
\begin{aligned}
F(x)^2
&=\left(\sum_{\alpha: a_\alpha>0 }
\frac{r_\alpha(x)}{\sqrt{a_\alpha}}\sqrt{a_\alpha}\right)^2\\
&\le\left(\sum_{\alpha: a_\alpha>0}\frac{r_\alpha(x)^2}{a_\alpha}\right)
\left( \sum_{\alpha: a_\alpha>0} a_\alpha \right)
=\Psi(x,a).
\end{aligned}
\label{eq-shot-bound}
\end{equation}}
Equality holds when the shot fractions are proportional to the group standard deviations:
\begin{equation}
a_\alpha^\star(x)=\frac{r_\alpha(x)}{F(x)},
\ \Psi(x,a^\star(x))=F(x)^2.
\label{eq-shot-elimination}
\end{equation}
When $r_\alpha(x)=0$, positive definiteness implies $x_\alpha=0$, so the zero fraction in Eq.~\eqref{eq-shot-elimination} respects the condition in Section~\ref{sec-estimator-variance}.

Eliminating the shot fractions therefore leaves the minimization of $F(x)^2$ over feasible coefficients.
Since squaring preserves the ordering of nonnegative values, it suffices to solve
\begin{equation}
F^\star=\min_{\sum_\alpha S_\alpha^Tx_\alpha=c}
\sum_\alpha\|\Gamma_\alpha^{1/2}x_\alpha\|_2,
\ \Phi_{\mathcal G}^\star=(F^\star)^2.
\label{eq-socp}
\end{equation}
Introduce one scalar $\tau_\alpha\in\mathbb R$ to bound each group norm and collect them into $\tau=(\tau_1,\ldots,\tau_L)^T\in\mathbb R^L$.
The SOCP is
\begin{equation}
\begin{aligned}
F^\star=\min_{x,\tau}\quad&\sum_{\alpha=1}^{L}\tau_\alpha\\
\text{subject to}\quad&\sum_{\alpha=1}^{L}S_\alpha^Tx_\alpha=c,\\
&\|\Gamma_\alpha^{1/2}x_\alpha\|_2\le \tau_\alpha,
\forall \alpha.
\end{aligned}
\label{eq-conic-primal}
\end{equation}
Each norm constraint places $(\tau_\alpha,\Gamma_\alpha^{1/2}x_\alpha)\in\mathbb R^{d_\alpha+1}$ in $\mathcal Q_{d_\alpha+1}$.
At an optimum, $\tau_\alpha=\|\Gamma_\alpha^{1/2}x_\alpha\|_2$ for every group, because any larger value could be reduced without changing the coefficients.
An optimal solution $(x^\star,\tau^\star)$ therefore recovers the shot fractions as
\begin{equation}
a_\alpha^\star
=\frac{\tau_\alpha^\star}{\sum_\beta \tau_\beta^\star}
=\frac{\|\Gamma_\alpha^{1/2}x_\alpha^\star\|_2}{F^\star}.
\label{eq-socp-map}
\end{equation}
The optimal objective value of SOCP $F^\star$ is the square root of the optimal model variance coefficient.
For a total budget of $M$ shots, the minimum variance predicted by the model is $(F^\star)^2/M$.
Thus the SOCP solves the joint problem in Section~\ref{sec-estimator-variance} under its continuous allocation assumption.
Appendix~\ref{app-socp-duality} proves that an optimum exists and gives the corresponding dual problem.


\subsection{Reduction to an allocation problem}\label{sec-coefficient-reduction}

We now eliminate the estimator coefficients from Eq.~\eqref{eq-joint} while retaining the shot fractions as the decision variables.
Fix an allocation $a\in\Delta_L$ such that every retained Pauli operator belongs to at least one group with $a_\alpha>0$.
Let $\mathcal A_+(a)=\{\alpha\in\{1,\ldots,L\}:a_\alpha>0\}$ index the groups receiving shots.
The coefficient problem at this allocation is
\begin{equation}
\begin{aligned}
\Phi(a)=\min_{\{x_\alpha\}_{\alpha\in \mathcal A_+(a)}}\quad&
\sum_{\alpha\in \mathcal A_+(a)}\frac{x_\alpha^T\Gamma_\alpha x_\alpha}{a_\alpha}\\
\text{subject to}\quad&\sum_{\alpha\in \mathcal A_+(a)}S_\alpha^Tx_\alpha=c,
\end{aligned}
\label{eq-coefficient-subproblem}
\end{equation}
with $x_\alpha=0$ for every group outside $\mathcal A_+(a)$.
We write $\Phi_{\mathcal G}(a)=\Phi(a)$ when the group set needs to be explicit.

Introduce the Lagrange multiplier vector $\mu\in\mathbb R^{N_P}$, with one entry for each unbiasedness equality in Eq.~\eqref{eq-unbiasedness}.
Scaling the objective by $1/2$ simplifies differentiation without changing the minimizing coefficients, giving the Lagrangian
\begin{equation}
\mathcal L_a(x,\mu)
=\frac12\sum_{\alpha\in \mathcal A_+(a)}\frac{x_\alpha^T\Gamma_\alpha x_\alpha}{a_\alpha}
-\mu^T\left(\sum_{\alpha\in \mathcal A_+(a)}S_\alpha^Tx_\alpha-c\right).
\label{eq-coefficient-lagrangian}
\end{equation}
For this quadratic problem with linear equality constraints, the Karush--Kuhn--Tucker (KKT) conditions require the coefficient derivatives of the Lagrangian to vanish and the unbiasedness constraints to hold~\cite{boyd2004convex}:
\begin{equation}
\begin{aligned}
\nabla_{x_\alpha}\mathcal L_a
&=\frac{\Gamma_\alpha x_\alpha}{a_\alpha}-S_\alpha\mu=0,
&&\alpha\in \mathcal A_+(a),\\
\sum_{\alpha\in \mathcal A_+(a)}S_\alpha^Tx_\alpha&=c.
\end{aligned}
\label{eq-coefficient-stationarity}
\end{equation}
The first row is a vector equation in $\mathbb R^{d_\alpha}$ for each measured group, and the second is a vector equation in $\mathbb R^{N_P}$.
The fractions $a_\alpha$ are fixed in this subproblem, so their nonnegativity constraints introduce no additional KKT conditions here.
The first row gives $x_\alpha^\star(a)=a_\alpha\Gamma_\alpha^{-1}S_\alpha\mu$.
Substituting this expression into the coefficient constraint gives
\begin{equation}
c=\sum_{\alpha\in \mathcal A_+(a)}S_\alpha^Tx_\alpha^\star(a)
=\left(\sum_{\alpha\in \mathcal A_+(a)}a_\alpha S_\alpha^T\Gamma_\alpha^{-1}S_\alpha\right)\mu.
\label{eq-coefficient-substitution}
\end{equation}
This identity leads to the embedded group precision matrices $W_\alpha\in\mathbb R^{N_P\times N_P}$ and their sum weighted by the allocation fractions $K(a)\in\mathbb R^{N_P\times N_P}$,
\begin{equation}
W_\alpha=S_\alpha^T\Gamma_\alpha^{-1}S_\alpha,
\quad K(a)=\sum_\alpha a_\alpha W_\alpha.
\label{eq-precision}
\end{equation}
In the eigenbasis of $\Gamma_\alpha$, the inverse covariance weights each direction by the reciprocal of its model variance.
The factor $a_\alpha$ accounts for the sampling effort assigned to the group.
The selection matrices place the entries of $\Gamma_\alpha^{-1}$ in the rows and columns indexed by $G_\alpha$, with zeros elsewhere in $W_\alpha$.
Thus $K(a)$ combines the inverse covariance matrices of all groups according to their shot fractions.

Every retained Pauli operator belongs to at least one group receiving shots, and each $\Gamma_\alpha$ is positive definite.
These conditions make $K(a)$ positive definite, as shown in Appendix~\ref{app-equivalence}.
Solving the KKT system therefore reduces to
\begin{equation}
K(a)\mu=c,
\quad x_\alpha^\star(a)=a_\alpha\Gamma_\alpha^{-1}S_\alpha\mu.
\label{eq-assignment}
\end{equation}
This formula also assigns zero coefficients to unused groups.
The multiplier depends on the allocation through $\mu=\mu(a)=K(a)^{-1}c\in\mathbb R^{N_P}$.
The objective in Eq.~\eqref{eq-coefficient-subproblem} is strictly convex in the coefficients of groups receiving shots, so the KKT system gives the unique minimizing coefficient assignment at the fixed allocation.

Substitution into the unscaled objective in Section~\ref{sec-estimator-variance} now gives
\begin{equation}
\begin{aligned}
\Phi(a)=\Psi(x^\star(a),a)
&=\sum_{\alpha\in \mathcal A_+(a)}a_\alpha\mu^TW_\alpha\mu\\
&=\mu^TK(a)\mu
=c^T\mu
=c^TK(a)^{-1}c.
\end{aligned}
\label{eq-reduced-variance}
\end{equation}
Equation~\eqref{eq-reduced-variance} is the objective value after solving the coefficient KKT system.
Minimizing this value over the shot fractions gives the reduced problem
\begin{equation}
\Phi_{\mathcal G}^\star=\min_{a\in\Delta_L}\Phi(a),
\
\Phi(a)=\begin{cases}
c^TK(a)^{-1}c,&K(a)\succ0,\\
+\infty,&K(a)\text{ singular}.
\end{cases}
\label{eq-allocation-problem}
\end{equation}
Because $S_\alpha$ selects the entries belonging to group $\alpha$, a singular $K(a)$ means that at least one Pauli operator with $c_j\ne0$ belongs to no group receiving shots.
Such an allocation cannot satisfy the coefficient constraint, which explains the infinite objective value.

To establish convexity, introduce an auxiliary vector $\xi\in\mathbb R^{N_P}$, with one component for each coefficient equality.
The reduced objective has the representation
\begin{equation}
\Phi(a)=\sup_{\xi\in\mathbb R^{N_P}}
\left(2c^T\xi-\sum_\alpha a_\alpha \xi^TW_\alpha \xi\right).
\label{eq-variational-variance}
\end{equation}
For $K(a)\succ0$, completing the square gives the maximum at $\xi=K(a)^{-1}c=\mu$.
For a singular $K(a)$, some Pauli operator is unmeasured, so the term $2c_j\xi_j$ is unbounded above without a corresponding quadratic term.
For each fixed $\xi$, the expression in parentheses is linear in $a$. Taking its supremum therefore gives a convex objective on $\Delta_L$.
This gives a convex optimization over allocations, although the optimal allocation need not be unique.
Both eliminations preserve the joint optimum, with
\begin{equation}
\min_{a\in\Delta_L}\Phi(a)
=\min_{\sum_\alpha S_\alpha^Tx_\alpha=c}F(x)^2
=(F^\star)^2.
\label{eq-two-reductions}
\end{equation}

Let $\pi_j(a)=\sum_{\alpha:j\in G_\alpha}a_\alpha\in[0,1]$ denote the fraction of shots measuring term $j$.
For identity group covariances $\Gamma_\alpha=I_{d_\alpha}$, $\mu_j=c_j/\pi_j(a)$ and $x_{\alpha j}^\star=a_\alpha c_j/\pi_j(a)$, so the contribution of each batch is proportional to its number of samples of that operator.
With correlated outcomes, the system $K(a)\mu=c$ also accounts for covariance contributions across jointly measured terms.
The multiplier vector $\mu$ from the coefficient KKT system determines the group scores below.

\subsection{Optimal shot allocation}\label{sec-allocation-optimization}

The allocation objective in Eq.~\eqref{eq-allocation-problem} lets us compare the variance changes produced by assigning shots to different groups.
Using $\partial K/\partial a_\alpha=W_\alpha$ and the inverse derivative identity $dK^{-1}=-K^{-1}(dK)K^{-1}$ gives
\begin{equation}
\frac{\partial\Phi}{\partial a_\alpha}
=-c^TK(a)^{-1}W_\alpha K(a)^{-1}c
=-\mu^TW_\alpha\mu
=-\omega_\alpha(a).
\label{eq-gradient}
\end{equation}
Here $\omega_\alpha(a)=\mu^TW_\alpha\mu \ge0$ is the scalar group score.
The derivative already accounts for the optimal coefficient assignment at each allocation.
It measures the decrease in $\Phi$ when one fraction increases with the others held fixed.
A feasible change must preserve the total shot budget, so shots must instead be transferred between groups.
Moving a small fraction from a used group with a lower score to a group with a higher score decreases the model variance to first order.

The mean score weighted by the shot fractions is $\Phi(a)$.
For optimization over $a$, the KKT conditions include the nonnegativity and normalization of the shot fractions.
After eliminating their Lagrange multipliers, these conditions state that every group receiving shots has score $\Phi_{\mathcal G}^\star$, while every group with zero shots has no larger score:
\begin{equation}
\begin{cases}
\omega_\alpha(a^\star) = \Phi_{\mathcal{G}}^{\star} & \quad\text{if }a_{\alpha}^{\star} > 0, \\
\omega_\alpha(a^\star) \leq \Phi_{\mathcal{G}}^{\star} & \quad\text{if }a_{\alpha}^{\star} = 0.
\end{cases}\label{eq-kkt}
\end{equation}
These conditions equalize the first-order variance benefit of shots among used groups, without requiring equal shot counts.
A group can become unused when other groups supply its contribution more effectively.
Convexity makes Eq.~\eqref{eq-kkt} sufficient for optimality on the specified group set.
Appendix~\ref{app-allocation-kkt} derives the KKT conditions for the shot fractions.

Algorithm~\ref{alg-optimization} applies a Frank--Wolfe update to the reduced allocation problem.
Starting from an allocation in which every Pauli operator belongs to a group receiving shots, it transfers a fraction of the current allocation to a group with the highest score.
For the selected group index $q\in\{1,\ldots,L\}$, let $e_q\in\mathbb R^L$ be the $q$th standard basis vector: $(e_q)_\alpha=1$ if $\alpha=q$ and zero otherwise.
It is the vertex of $\Delta_L$ that assigns all shots to group $q$.
The step size $\gamma\in[0,1]$ minimizes $\Phi((1-\gamma)a+\gamma e_q)$ by varying this single parameter, a procedure called a line search.
The trial parameter is denoted by $\eta$ in Algorithm~\ref{alg-optimization}, and the coefficients are optimal at each trial allocation.
\textcolor{black}{
Algorithm~\ref{alg-optimization} uses the prescribed dimensionless relative tolerance $\varepsilon_{\mathrm{rel}}=10^{-4}$ in its stopping test.}

\begin{algorithm}[ht]
\caption{Coefficient and shot optimization}\label{alg-optimization}
\small
\begin{algorithmic}[1]
\Require Hamiltonian coefficients $c$, QWC groups $\mathcal G=\{G_\alpha\}_{\alpha=1}^{L}$, positive-definite covariance matrices $\{\Gamma_\alpha\}$, initial allocation $a^{(0)}\in\Delta_L$ with $K(a^{(0)})\succ0$
\Ensure Allocation $a$, coefficients $x$, vector $\mu$ satisfying $K(a)\mu=c$, model variance coefficient $\Phi$
\State $W_\alpha\gets S_\alpha^T\Gamma_\alpha^{-1}S_\alpha\quad(\alpha=1,\ldots,L)$
\State $a\gets a^{(0)}$
\Loop
    \State $K\gets\sum_{\alpha=1}^{L}a_\alpha W_\alpha$
    \State Solve $K\mu=c$ and set $\Phi\gets c^T\mu$
    \State Choose $q\in\arg\max_{\alpha\in\{1,\ldots,L\}}\mu^TW_\alpha\mu$
    \If{$\mu^TW_q\mu \le (1+\varepsilon_{\mathrm{rel}})\Phi$}
        \State \textbf{break}
    \EndIf
    \State Choose $\gamma\in\arg\min_{\eta\in[0,1]}\Phi((1-\eta)a+\eta e_q)$
    \State $a\gets(1-\gamma)a+\gamma e_q$
\EndLoop
\State $x_\alpha\gets a_\alpha\Gamma_\alpha^{-1}S_\alpha\mu\quad(\alpha=1,\ldots,L)$
\State \Return $(a,x,\mu,\Phi)$
\end{algorithmic}
\end{algorithm}
\textcolor{black}{
For an initial measurement group set, the underlying Frank--Wolfe iteration in Algorithm~\ref{alg-optimization} attains an $O(1/t)$ objective error bound under exact coefficient recovery and line minimization.
Appendix~\ref{app-convergence} establishes the required curvature bound by showing that $K(a)$ remains uniformly positive definite along the iterates.
Convexity also gives $0\le\Phi(a)-\Phi_{\mathcal G}^\star\le g(a)$, where $g(a)=\max_\alpha\omega_\alpha(a)-\Phi(a)$.
With exact coefficient recovery, this gap also assesses allocations returned by other solvers.
Numerical calculations used L-BFGS-B~\cite{byrd1995limited} on the parameterized allocation objective.
This quasi-Newton method approximates curvature from changes in the iterates and gradients, with coefficients recovered through Eq.~\eqref{eq-assignment} at each evaluation.
The original allocation variables were assessed using the primal--dual gap and the feasibility and residual checks specified in Appendix~\ref{app-protocol-numerics}.}

\subsection{Scoring candidate groups}\label{sec-certificates}

Let $G\subseteq\{1,\ldots,N_P\}$ be a candidate measurement group that is not in the current set $\mathcal G$.
Define its selection matrix $S_G\in\{0,1\}^{|G|\times N_P}$ and positive-definite model covariance $\Gamma_G\in\mathbb R^{|G|\times|G|}$ as for the existing groups.
The vector $\mu_G=S_G\mu\in\mathbb R^{|G|}$ selects the entries of the current multiplier vector corresponding to the candidate's Pauli operators.
The scalar candidate score is
\begin{equation}
\omega_G(a) = \mu_{G}^{T}\Gamma_{G}^{- 1}\mu_{G}.\label{eq-candidate-score}
\end{equation}
We also write $W_G=S_G^T\Gamma_G^{-1}S_G\in\mathbb R^{N_P\times N_P}$ for the corresponding matrix with rows and columns indexed by all $N_P$ Pauli operators.

For identity covariances, $\omega_G(a)=\sum_{j\in G}c_j^2/\pi_j(a)^2$.
At equal coefficient magnitudes, a less frequently measured term therefore contributes more to the score.
With correlated outcomes, $\Gamma_G^{-1}$ also accounts for how the candidate's outcomes can be combined.
The score changes when reoptimization changes $\mu$, even though the Hamiltonian and covariance model remain fixed.

For the enlarged group set $\mathcal G\cup\{G\}$, let $\Phi_{\mathcal G\cup\{G\}}$ denote the reduced objective in Eq.~\eqref{eq-allocation-problem}.
Write the old allocation in this set as $\widetilde a=(a^T,0)^T\in\mathbb R^{L+1}$ and let $e_G=e_{L+1}\in\mathbb R^{L+1}$ have entry one in the new group's position and zeros elsewhere.
Here the standard basis vector belongs to the enlarged allocation space, whose last coordinate represents $G$.
Assigning a fraction $\gamma\in[0,1]$ to the candidate gives $(1-\gamma)\widetilde a+\gamma e_G\in\Delta_{L+1}$.
This transfer preserves the total shot budget and has initial derivative
\begin{equation}
\left. \frac{d\Phi_{\mathcal G\cup\{G\}}((1-\gamma)\widetilde a+\gamma e_G)}{d\gamma} \right|_{\gamma=0^+}
=\Phi(a)-\omega_G(a).\label{eq-direction}
\end{equation}
The term $\Phi(a)$ is the first-order cost of reducing measurements in the old groups, and $\omega_G(a)$ is the candidate's first-order benefit.
Coefficients are reassigned optimally as the allocation changes.

At an exact optimum of the old set of groups, $\omega_G(a^\star)>\Phi_{\mathcal G}^\star$ guarantees a variance decrease for a sufficiently small transfer.
If the inequality fails, the old allocation satisfies all optimality conditions for the enlarged problem, so adding this candidate cannot lower its optimum.
Appendix~\ref{app-equivalence} proves the criterion and its extension to a finite candidate set.

Including all Hamiltonian terms measurable in the existing bases does not exhaust the possible combinations supplied by other bases.
A new basis can therefore satisfy the improvement criterion even after complete inclusion of compatible terms and joint estimator optimization on the old set.
As shots move to the candidate, $\mu$ and the marginal scores change.
We therefore use the current scores to rank candidate groups and reoptimize the estimator to determine the reduction achieved by an accepted addition.

\section{Construction of QWC measurement groups}\label{sec-construction}

We use the score in Eq.~\eqref{eq-candidate-score} to construct additional QWC groups from the current optimized estimator.
Starting from a single Pauli term, we add compatible terms according to their score increments.
Each nonidentity factor fixes a local measurement axis, and subsequent additions must respect the axes already fixed.
Starting from different terms can produce different compatible combinations, and reoptimization after each accepted group updates the estimator used for the next construction step.

\subsection{Growing candidate groups}

The value of adding a group depends on the group already constructed.
With diagonal covariances, its score contribution is independent of the other members.
With correlated measurement results, even a term with a small score when measured alone can have a large increment because of its covariances with those members.
We therefore evaluate each compatible addition relative to the group constructed so far.
All scores in this subsection use the current optimized allocation, and its multiplier vector $\mu$ is held fixed while a candidate is constructed.

Let $G$ be the group constructed so far and $i\notin G$ the index of a Pauli operator compatible with every operator in $G$.
The vector $u\in\mathbb R^{|G|}$ contains the model covariances between the outcome of $P_i$ and the outcomes of the operators in $G$, in the ordering used for $\Gamma_G$.
The scalar $\Gamma_{i i}$ is the model variance of the $P_i$ outcome.
We require the covariance assigned to a given pair of Pauli outcomes to remain unchanged when other terms are added to or removed from a group.
Thus the enlarged covariance matrix $\Gamma_{G\cup \{i \}} \in \mathbb R^{(|G|+1) \times (|G|+1)}$ contains $\Gamma_G$ as the submatrix for the original terms.
The covariance model remains fixed during construction, and all group covariance matrices are positive definite.

For outcomes with their means subtracted, $\Gamma_G^{-1}u\in\mathbb R^{|G|}$ gives the coefficients for linearly predicting the outcome of $P_i$ from the outcomes in $G$.
These coefficients minimize the mean squared prediction error under the covariance model.
The residual, defined as that outcome minus its linear prediction, has variance $\Gamma_{ii}-u^T\Gamma_G^{-1}u$.
Let $\mu_i$ denote the entry of $\mu$ associated with $P_i$.
Applying the block inverse identity to the enlarged covariance matrix gives the change in group score,
\begin{equation}
\Delta(i|G) = \omega_{G \cup \left\{ i \right\}} - \omega_G = \frac{\left( \mu_{i} - u^{T}\Gamma_{G}^{- 1}\mu_{G} \right)^{2}}{\Gamma_{ii} - u^{T}\Gamma_{G}^{- 1}u} \geq 0.\label{eq-increment}
\end{equation}
The numerator applies the same linear subtraction to the entries of $\mu$, and the denominator is strictly positive under the covariance assumptions.
A large increment requires a large squared residual component of $\mu$ relative to the residual variance.
A small denominator alone is insufficient because the covariances between $P_i$ and the terms in $G$ affect both quantities.
Interpreting the denominator as a conditional variance requires additional assumptions.

At fixed $\mu$, we repeatedly add the compatible term with the largest increment until no further term can be included.
Since $\omega_{G}$ is common to all compatible additions at a given step,  this rule also maximizes the score of the enlarged group at that step.
We retain additions with zero increments so that the resulting group is maximal. No further retained Pauli operator can then be included while preserving QWC.
Any unspecified local axes can then be assigned to complete a measurement basis.
No omitted Hamiltonian term is measurable in that basis, since such a term would already have been compatible with the group.
Choosing the largest current increment can exclude terms belonging to another useful group, so this rule need not construct the group with the largest possible score.

\subsection{Expansion of the group set}

\begin{algorithm}[t]
\caption{Variance guided measurement group construction (VGMC)}
\label{alg-expansion}
\small
\begin{algorithmic}[1]
\Require Hamiltonian $H$; covariance model $\{\Gamma_G\}$;
initial QWC groups $\mathcal G_0$; addition budget $B$
\Ensure QWC groups $\mathcal G$; shot fractions $a$;
estimator coefficients $x$; variance coefficient $\Phi$
\State $\mathcal G\gets\mathcal G_0$
\State $(a,x,\mu,\Phi)\gets
\Call{Optimize}{\mathcal G,\mathbf1_{L_0}/L_0,\varepsilon_{\mathrm{rel}}}$
\Comment{Algorithm~\ref{alg-optimization}}
\For{$t=1,\ldots,B$}
    \State For each $j\in[N_P]$, grow $G_j$ from $\{j\}$
    by repeatedly adding a term maximizing $\Delta(i\mid G_j)$
    over $\mathcal I(G_j)$ until $\mathcal I(G_j)=\varnothing$
    [Eq.~\eqref{eq-increment}]
    \State $\mathcal C\gets\{G_j:j\in[N_P]\}\setminus\mathcal G$
    \State \textbf{if} $\mathcal C=\varnothing$ \textbf{then break}
    \State Choose $G^\star \in \arg\max_{G \in \mathcal C}\omega_G$
    \Comment{Eq.~\eqref{eq-candidate-score}}
    \State \textbf{if} $\omega_{G^\star}\leq \left(1+\varepsilon_{\mathrm{adm}} \right) \Phi$ \textbf{then break}
    \Comment{Eq.~\eqref{eq-direction}}
    \State $\mathcal G\gets\mathcal G\cup\{G^\star\}$
    \State $(a,x,\mu,\Phi)\gets
    \Call{Optimize}{\mathcal G,(a^T,0)^T, \varepsilon_{\mathrm{rel}}}$
\EndFor
\State \Return $(\mathcal G,a,x,\Phi)$
\end{algorithmic}
\end{algorithm}

We initialize the design with overlapped SI, include every Hamiltonian term measurable in its selected bases, and merge groups with identical Hamiltonian term index sets.
We denote this initial group set by $\mathcal G_0$ and jointly optimize its estimator coefficients and shot fractions.
Let $\mu_0\in\mathbb R^{N_P}$ be the multiplier vector for this optimized initial design.
With exact optimization, this baseline already incorporates the gains obtainable solely by including omitted compatible terms or reallocating coefficients and shots within the original bases.

At each expansion step, we use the current optimized $\mu$ to construct a maximal QWC group from every Hamiltonian term.
After removing duplicates and groups already present, we select a candidate with the largest score.
At an exact optimum of the current design, a score exceeding $\Phi_{\mathcal G}^\star$ guarantees a strict decrease in the optimum after adding the candidate and reoptimizing, by Eq.~\eqref{eq-direction}.
If no candidate exceeds this threshold, construction stops for the generated candidate set.

Algorithm~\ref{alg-expansion} allows at most $B$ accepted additions, where $B$ is a nonnegative integer.
The budget counts accepted groups even if reoptimization later assigns them zero shots.
Write $[N_P]=\{1,\ldots,N_P\}$ and let $ \mathcal I(G)=\{i\in[N_P] \setminus G:G\cup\{i\}\text{ is QWC}\}$ list the terms that can be added while preserving QWC.
For candidates $G^{(j)}$ constructed from each $P_j$, the set $ \mathcal C=\{G^{(j)}:j\in [N_P]\}\setminus\mathcal G$ contains the distinct groups not already included.

Let $L_0=| \mathcal G_0|$ and let $\mathbf1_{L_0}\in\mathbb R^{L_0}$ have every entry equal to one.
The initial allocation $\mathbf1_{L_0}/L_0$ assigns the same shot fraction to each initial group.
When a group is added to a set of $L$ groups, the column vector $(a^T,0)^T \in \mathbb R^{L+1}$ retains the old fractions and assigns zero to the new group as a starting point for re-optimization.
The old groups still measure every retained Pauli operator, so this starting allocation remains feasible.
The dimensions of the allocation and coefficient vectors change with the group set.
\textcolor{black}{
Algorithm~\ref{alg-expansion} passes the relative tolerance $\varepsilon_{\mathrm{rel}}$ to each allocation solve.
A candidate is admitted when $\omega_G>(1+\varepsilon_{\mathrm{adm}})\Phi$, where $\varepsilon_{\mathrm{adm}} = 10^{-6}$ is the numerical comparison margin.}

The numerical calculations used batches to distribute the reoptimization cost across several additions.
A batch keeps the current $\mu$ fixed while seeds are processed in waves, deduplicates the resulting groups, and selects the candidates with the highest scores above this threshold.
Scanning terminates once the batch reaches capacity with a sufficient number of candidates or once all seeds have been processed. 
The coefficients and shot fractions are then reoptimized on the enlarged group set before the next search.

\subsection{Quantifying improvement and assessing optimality}

Let $\mathcal G_t$ be the group set after $t$ accepted additions and let $G_t^{\mathrm{new}}$ be the group accepted at step $t\ge1$.
The dimensionless relative decrease in the optimized variance coefficient is
\begin{equation}
D_t=1-\frac{\Phi_{\mathcal G_t}^\star}{\Phi_{\mathcal G_{t-1}}^\star},
\ \mathcal G_t=\mathcal G_{t-1}\cup\{G_t^{\mathrm{new}}\}.
\label{eq-single-gain}
\end{equation}
Both optima use the same covariance model.
When that model matches the measured state, $D_t$ is also the relative reduction in estimator variance at a fixed shot budget under continuous allocation.
With approximate covariances, it measures the reduction predicted by the model.
The variance on the measured state must be evaluated separately.
Reoptimization may assign zero shots to previously used groups, so the total number of available groups and the number of groups that receive shots are distinct quantities.

With exact coefficient and shot optimization, failure to find a candidate with score above $\Phi_{\mathcal G}^\star$ establishes optimality over the current group set together with the candidates generated in that round.
It does not establish optimality over all QWC groups, because the construction rule need not generate every improving group.
If the addition budget is exhausted, construction stops without testing whether further candidates improve the design.
If no group is accepted, the optimized initial estimator remains the final design and the additional variance reduction is zero.

\subsection{Computational cost}\label{sec-construction-cost}

We count the dense linear algebra operations in VGMC with covariance entries and QWC compatibility available.
Let $d_{\max}\le N_P$ be the largest number of terms in any generated or retained group, and let $L_{\max}=L_0+B$ bound the number of retained groups.
The current implementation evaluates each extension of a group containing $k$ terms by factorizing its $k\times k$ covariance matrix, at a cost of $O(k^3)$.
Scanning at most $N_P$ extensions at each growth step gives $O(N_Pd_{\max}^4)$ per seed and $O(N_P^2d_{\max}^4)$ per complete construction round.

Let $E_{\mathrm{tot}}$ be the total number of allocation evaluations across the initial optimization and all reoptimizations, including trials during line searches and final optimality checks.
Preparing the group covariance factors and inverse blocks costs $O(L_{\max}d_{\max}^3)$ per optimization call, and these matrices are reused within that call.
Each allocation evaluation assembles $K(a)$, solves $K(a)\mu=c$, and scores all retained groups at a cost of $O(N_P^3+L_{\max}d_{\max}^2)$.
Final coefficient recovery costs $O(L_{\max}d_{\max}^2)$ and is covered by the same bound.
Algorithm~\ref{alg-expansion} performs at most $B$ construction rounds and $B+1$ optimization calls.
Combining candidate construction, matrix preparation, and allocation optimization gives the total dense arithmetic cost
$O\Bigl(
B N_P^2d_{\max}^4
+(B+1)L_{\max}d_{\max}^3
+E_{\mathrm{tot}}\bigl(N_P^3+L_{\max}d_{\max}^2\bigr)
\Bigr).$

\section{Experiments}\label{sec-experiments}

\begin{table*}[t]
\caption{Estimator variance on the reference states $V$ (Hartree$^2$).
$N_e$ is the electron count, $N_q$ the qubit count, $N_P$ the number of nonidentity Pauli terms, and $L_0$ the number of completed initial groups.
Boldface marks the lowest variance coefficient in each row, including ties at the displayed precision.
}\label{tab-molecular-variance}
\begin{ruledtabular}
\begin{tabular}{lccccrrrrrr}
\textrm{Molecule} &
$N_e$ &
$N_q$ &
$N_P$ &
$L_0$ &
\multicolumn{1}{c}{\textrm{LF}~\cite{verteletskyi2020minimumclique}} &
\multicolumn{1}{c}{\textrm{OGM}~\cite{wu2023ogm}} &
\multicolumn{1}{c}{\textrm{SI}~\cite{crawford2021efficient}} &
\multicolumn{1}{c}{\textrm{ICS}~\cite{yen2023deterministic}} &
\multicolumn{1}{c}{\textrm{SOCP}~\cite{sagnol2011multiresponse}} &
\multicolumn{1}{c}{\textrm{VGMC}} \\
\colrule
\multicolumn{11}{l}{\textit{Design using exact covariances}} \\
H$_2$  & 2  & 4  & 14   & 3    & \textbf{0.1364} & 0.1756             & \textbf{0.1364} & \textbf{0.1364} & \textbf{0.1364} & \textbf{0.1364} \\
LiH    & 4  & 12 & 630  & 155  & 5.055           & 3.592              & 2.094           & 0.9810          & 0.9748          & \textbf{0.7843} \\
BeH$_2$ & 6 & 14 & 665  & 183  & 14.59           & 11.71              & 6.344           & 4.291           & 4.288           & \textbf{3.508}  \\
H$_2$O & 10 & 14 & 1085 & 334  & 125.0           & 78.60              & 48.61           & 13.61           & 13.45           & \textbf{10.46}  \\
NH$_3$ & 10 & 16 & 3608 & 1359 & 388.0           & 148.9   & 97.00           & 45.67           & 44.79           & \textbf{34.32}  \\
\colrule
\multicolumn{11}{l}{\textit{Design using CISD covariances}} \\
H$_2$  & 2  & 4  & 14   & 3    & \textbf{0.1364} & 0.1756             & \textbf{0.1364} & \textbf{0.1364} & \textbf{0.1364} & \textbf{0.1364} \\
LiH    & 4  & 12 & 630  & 155  & 5.055           & 3.592              & 2.094           & 0.9812          & 0.9771          & \textbf{0.7859} \\
BeH$_2$ & 6 & 14 & 665  & 183  & 14.59           & 11.71              & 6.345           & 4.405           & 4.612           & \textbf{3.714}  \\
H$_2$O & 10 & 14 & 1085 & 334  & 125.0           & 78.60              & 48.61           & 15.67           & 14.29           & \textbf{11.36}  \\
NH$_3$ & 10 & 16 & 3608 & 1359 & 388.1           & 148.9   & 97.00           & 46.19           & 46.68           & \textbf{35.76}  \\
\end{tabular}
\end{ruledtabular}
\end{table*}

We evaluated whether group construction reduces variance in energy estimation beyond joint optimization on the initial measurement bases.
Molecular comparisons tested this question with exact and approximate covariance information. Random construction and growth ablations examined the contribution of candidate selection, and simulations with finite shot budgets tested the resulting designs near chemical accuracy.

\subsection{Molecular systems and evaluation}
\label{sec-molecular-systems}

We consider H$_2$, LiH, BeH$_2$, H$_2$O, and NH$_3$ in the complete STO-3G orbital spaces with the Bravyi--Kitaev mapping.
The H--H, Li--H, Be--H, O--H and N--H bond lengths are all 1~\AA.
BeH$_2$ is linear. The H--O--H and H--N--H bond angles in H$_2$O and NH$_3$ are $107.6^\circ$ and $107^\circ$, respectively \cite{yen2023deterministic}.
Table~\ref{tab-molecular-variance} lists the electron counts ($N_e$), the sizes of the qubit Hamiltonians ($N_q$ the number of qubits and $N_P$ the number of Pauli terms) and initial group sets ($L_0$).
For each molecule, the reference state had the lowest energy among states with $N_e$ electrons and equal alpha and beta populations.
CISD supplies an approximate covariance model for designing the measurements,
with the resulting estimator evaluated on the reference state under either model.
We evaluate designs by the reference state variance coefficient $V=M\operatorname{Var}(\widehat E)$ in Hartree$^2$.
Appendix~\ref{app-protocol-inputs} specifies which spin orbitals are occupied in the reference determinant used to construct the CISD subspace.

Using the conic formulation, the SOCP baseline jointly optimized coefficients and shot fractions on the completed initial bases obtained by overlapped SI. VGMC expanded that design and reoptimized both variables.
%
SI and largest-first graph coloring (LF) supplied disjoint partitions and allocated shots in proportion to their group standard deviations under the chosen design covariance source. 
ICS performed five updates of the coefficients and shot allocations on overlapping groups using the Tequila implementation~\cite{Kottmann_Tequila_A_platform_2020}.
Overlapped grouping measurement (OGM) estimates each Pauli expectation value by averaging the outcomes recorded for that operator across overlapping QWC groups. Its group construction and shot allocation depend solely on the Hamiltonian terms and coefficients. We therefore used the same OGM design in the exact and CISD covariance comparisons.


\subsection{Variance reduction with exact and approximate covariances}

For designs based on exact covariances, VGMC reduced variance by $18.19\%$--$23.38\%$ relative to SOCP for $\text{LiH}$, $\text{BeH}_2$, $\text{H}_2\text{O}$, and $\text{NH}_3$ (Table~\ref{tab-molecular-variance}).
For H$_2$, the initial bases covered all maximal QWC groups, so construction added no basis and gave no further reduction.
In the four systems that benefited, replacing the five ICS updates with SOCP reduced variance by $0.07\%$--$1.93\%$ on the initial groups.
The larger improvement after expansion identifies useful measurement combinations beyond the estimation freedom already optimized on those bases.

The same four molecules benefited when CISD covariances guided construction and estimator optimization.
On the reference states, VGMC reduced variance by $19.47\%$--$23.39\%$ relative to SOCP designs using the same approximate information.
The expanded CISD designs also improved on the initial designs based on exact covariances by $13.39\%$--$20.16\%$.
In these comparisons, better measurement combinations more than compensated for the reduced accuracy of the covariance information. Construction based on exact covariances gave the lowest final variances, so covariance accuracy continued to matter after expansion.

\textcolor{black}{
With CISD covariances, SOCP yielded higher variances than ICS for  $\text{BeH}_2$ and $\text{NH}_3$, suggesting that covariance errors can outweigh the modest gains from more complete optimization of coefficients and shot fractions on the initial groups. VGMC uses the same covariance-dependent optimization and also yields higher variances with CISD than with exact covariances. Its advantage over both ICS and SOCP nevertheless persists in all four systems where groups are added, indicating that the larger gains from additional measurement combinations outweigh the loss associated with approximate covariance information in these benchmarks.}

\subsection{Candidate selection and growth ablations}
\begin{figure}[htbp]
\centering\includegraphics[width=.85\linewidth]{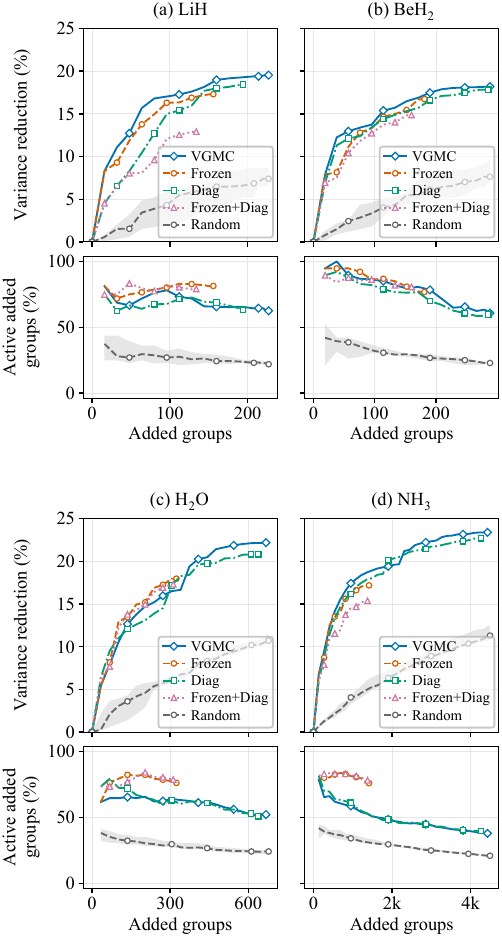}
\caption{Variance reduction and activity of added groups during measurement group construction.
Upper panels show the reduction relative to the optimized initial variance,
and lower panels show the percentage of added groups with $a_{\alpha}>10^{-6}$, where $a_{\alpha}$ is the shot allocation fraction.
All curves use designs based on exact covariances. Gray Random curves show the median over five independent random constructions.
}\label{fig-matched-groups}
\end{figure}
To determine whether the choice of additional groups mattered, we compared VGMC with five independent random compatible group constructions.
For each random construction, Hamiltonian terms were greedily combined in a random order to form a compatible basis. 
The corresponding group was completed with all measurable terms, and repeated groups were discarded. 
At each tested number of added groups, the expanded design was jointly reoptimized using the same initial group set.
At every tested nonzero addition count common to VGMC and the five independent random constructions, VGMC achieved lower variance on the reference state than all five controls (Fig.~\ref{fig-matched-groups}).
Matching addition counts holds the size of the expanded design fixed. The comparison therefore supports selecting groups with information from the optimized estimator.

We examined the information used in growth with three ablations. Frozen evaluated $\Delta(i\mid G)$ using the initial vector $\mu_0$, Diag omitted off-diagonal covariances from those increments, and Frozen+Diag combined both changes.
All variants used the current estimator and full covariance information to score completed candidates and jointly reoptimize the design.

At their respective stopping points, Diag retained approximately 94\%--98\% of VGMC's terminal reduction.
Much of the gain therefore remained with simpler increments when full covariances were retained for computing $\mu$, evaluating candidates, and optimizing the estimator.
Frozen+Diag had the smallest terminal reduction of the four guided variants in each system, showing the consequence of combining the two simplifications.

The lower panels show how reoptimization used the added groups. Frozen could retain a larger active fraction while achieving less variance reduction.
The quality of the estimator therefore depended on the combinations selected and their optimized contributions, rather than on the active fraction alone.

\subsection{Energy estimation with finite shots}
\begin{figure}[ht]
\centering\includegraphics[width=.8\linewidth]{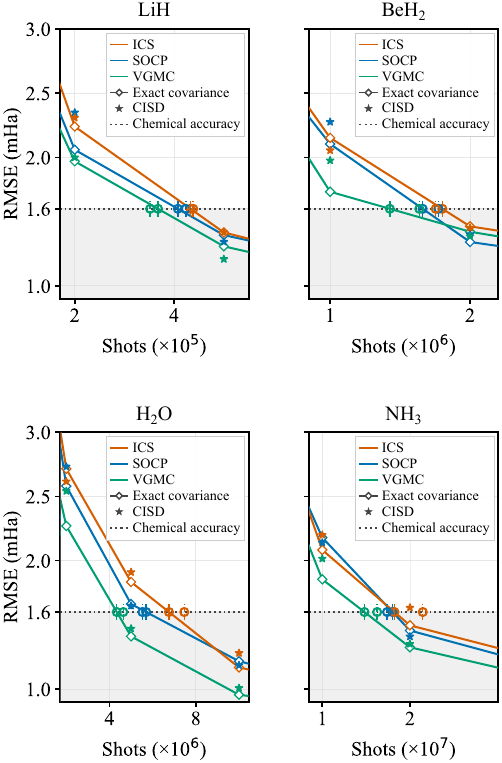}
\caption{Energy estimation error with finite shots near chemical accuracy.
The dotted line marks chemical accuracy (1.6 mHa), and the shaded region lies below this threshold.
Diamonds denote designs using exact covariances and stars denote CISD designs.
Each diamond or star gives the empirical RMSE of 300 independent estimates.
This reference is the lowest energy obtained by diagonalizing the molecular Hamiltonian in the complete STO-3G space, as detailed in~Appendix~\ref{app-protocol-inputs}.
Solid lines connect the diamonds.
The analytic RMSE discussed in the text is computed separately.
Hollow circles indicate approximate shot requirements obtained by interpolating threshold crossings between sampled budgets.
}\label{fig-energy-rmse}
\end{figure}

We converted the optimized ICS, SOCP, and VGMC shot fractions to integer allocations while holding their group sets and estimator coefficients fixed.
Each group with a nonzero coefficient vector first received one shot. The remaining budget was distributed in proportion to the optimized fractions by the largest-remainder rule, preserving the total budget and ensuring that every contributing group was sampled.
Sampling the joint measurement outcomes gave the empirical root mean square error (RMSE) shown in Fig.~\ref{fig-energy-rmse}. Each plotted RMSE was calculated from 300 independent energy estimates.
The analytic RMSE was calculated separately from the estimator variance under the integer allocation.


Near chemical accuracy ($1.6$~mHa), the empirical RMSE generally followed the ordering predicted for the integer allocations, with VGMC retaining an advantage over the initial designs.
Interpolated threshold crossings indicated shot savings of $9.8\%$--$23.9\%$ relative to SOCP for designs based on exact covariances and $6.5\%$--$16.9\%$ for CISD designs.
These crossing estimates connect the variance reductions to lower shot requirements at a specified energy precision.

\section{Conclusion}\label{sec-conclusion}
We have shown that an optimized energy estimator can guide the construction of additional measurement groups. Under a fixed positive definite design covariance model, the optimality conditions for coefficient assignment and shot allocation provide a criterion for identifying improving candidates. VGMC uses the associated score increments to construct QWC groups and reoptimizes the estimator after each accepted addition. The optimized estimator therefore determines how the available groups are used and provides guidance for extending the group set.

The molecular benchmarks demonstrate variance reductions beyond joint optimization on the initial measurement bases. 
Comparisons at matched addition counts support the contribution of estimator guided selection of new measurement combinations. In these considered molecular systems, CISD guided construction also gives lower reference state variance than the initial designs optimized with exact covariances. Thus, improved measurement combinations can compensate for approximate covariance information, although exact covariance construction still gives the lowest variances. Finite-shot simulations further indicate that these improvements reduce the estimated shot requirements near chemical accuracy.

The improvement criterion is exact for evaluated candidates under the stated model, while greedy construction provides a heuristic search for those candidates. Together, the exact criterion and heuristic search connect estimator optimization to the selection of local Pauli measurements, offering a route to reducing the sampling overhead of variational quantum algorithms.

\section{Acknowledgments}
We gratefully acknowledge support from the National Natural Science Foundation of China under Grant 62304037, the Jiangsu Provincial Scientific Research Center of Applied Mathematics under Grant BK20233002, the Start-up Research Fund of Southeast University under Grant RF1028623173, the Jiangsu Key Laboratory for Undersea Communications and Sensing under Grant 8506006260C, and the Big Data Computing Center of Southeast University and the Center for Fundamental and Interdisciplinary Sciences of Southeast University.

\section{Data Availability}
The data and code supporting this study are publicly available~\cite{vgmc_2026_v1}.

\newpage
\appendix
\section{Coefficient assignment and group scores}\label{app-equivalence}

\subsection{Zero shot fractions and coefficient assignment}
\label{app-coefficient-assign}
For each group, let $x_\alpha\in\mathbb R^{d_\alpha}$ and $a_\alpha\ge0$.
Extend the variance objective to zero shot fractions by defining
\begin{equation}
\begin{aligned}
q_\alpha(x_\alpha,a_\alpha) &=
\begin{cases}
x_\alpha^T\Gamma_\alpha x_\alpha/a_\alpha,&a_\alpha>0,\\
0,&a_\alpha=0\text{ and }x_\alpha=0,\\
+\infty,&a_\alpha=0\text{ and }x_\alpha\ne0,
\end{cases}
\\
\Psi(x,a)&=\sum_\alpha q_\alpha(x_\alpha,a_\alpha).
\end{aligned}
\label{eq-perspective}
\end{equation}
This definition requires zero coefficients in any group that receives no shots and makes $\Psi$ jointly convex in $x$ and $a$.

\begin{proposition}[Coefficient assignment]\label{thm-reduction}
Under the assumptions of Section~\ref{sec-model}, every allocation with $K(a)\succ0$ has the unique optimal coefficients in Eq.~\eqref{eq-assignment}.
With $\Phi(a)=+\infty$ when $K(a)$ is singular, minimizing $\Phi$ on $\Delta_L$ is a convex problem that attains the joint minimum in Eq.~\eqref{eq-joint}.
\end{proposition}

\begin{proof}
An allocation permits an unbiased estimator only if each retained Pauli operator belongs to at least one group with $a_\alpha>0$, since $c_j\ne0$ for every retained term.
For any nonzero $z\in\mathbb R^{N_P}$, at least one such group has $S_\alpha z\ne0$.
Positive definiteness of its covariance matrix therefore gives
\begin{equation}
z^TK(a)z=\sum_{\alpha:a_\alpha>0}
a_\alpha(S_\alpha z)^T\Gamma_\alpha^{-1}(S_\alpha z)>0.
\end{equation}
Conversely, if $K(a)$ is singular, some retained Pauli operator belongs to no group with positive shot fraction, so its coefficient equality cannot be satisfied.
For an allocation satisfying these conditions, the coefficient KKT system in Eq.~\eqref{eq-coefficient-stationarity} gives the assignment in Eq.~\eqref{eq-assignment}.

For any other feasible assignment, write $x=x^\star+\Delta x$, where $\Delta x\in\mathbb R^{N_x}$ has group components $\Delta x_\alpha\in\mathbb R^{d_\alpha}$.
Both assignments satisfy the coefficient equalities, so $\sum_\alpha S_\alpha^T\Delta x_\alpha=0$ and the cross term vanishes.
Thus
\begin{equation}
\Psi(x,a)=\Phi(a)+\sum_{\alpha:a_\alpha>0}
\frac{(\Delta x_\alpha)^T\Gamma_\alpha\Delta x_\alpha}{a_\alpha}.
\label{eq-excess}
\end{equation}
Positive definiteness proves uniqueness at fixed allocation.
Equation~\eqref{eq-variational-variance} establishes convexity of the reduced objective, and Appendix~\ref{app-socp-duality} establishes existence of an optimum.
\end{proof}

\subsection{Existence of an optimum and conic duality}\label{app-socp-duality}

We now show that the coefficient problem in Eq.~\eqref{eq-socp} has an optimal solution and identify its conic dual.
The feasible coefficient set is nonempty and closed.
Let $\lambda_{\min}$ denote the smallest eigenvalue and set $\sigma_\ast=\min_\alpha\lambda_{\min}(\Gamma_\alpha)>0$.
For the stacked coefficient vector $x\in\mathbb R^{N_x}$, the objective satisfies $F(x)\ge\sqrt{\sigma_\ast}\|x\|_2$.
The objective therefore tends to infinity as $\|x\|_2$ tends to infinity and hence attains a minimum on the feasible set.
Since $c\ne0$, this minimum is positive.
The recovery formulas in Eqs.~\eqref{eq-socp-map} and \eqref{eq-assignment} give optimal shot fractions and coefficients for the joint problem.

The dual problem uses an auxiliary vector $y\in\mathbb R^{N_P}$ to maximize $c^Ty$ subject to $y^TW_\alpha y\le1$ for every group, as in the conic formulation of multiresponse optimal design~\cite{sagnol2011multiresponse}.
Every feasible $y$ gives a lower bound $c^Ty\le F^\star$.
Because every retained term belongs to at least one available group, the coefficient equalities have a solution.
Choosing $\tau_\alpha>\|\Gamma_\alpha^{1/2}x_\alpha\|_2$ then satisfies every cone inequality strictly, so the primal and dual optimal objective values coincide.
For any optimal allocation $a^\star$, define $\mu^\star=K(a^\star)^{-1}c\in\mathbb R^{N_P}$.
The allocation conditions in Eq.~\eqref{eq-kkt} show that $y=\mu^\star/\sqrt{\Phi_{\mathcal G}^\star}$ attains the dual optimum $F^\star$.

\subsection{Allocation optimality and additional measurement groups}\label{app-allocation-kkt}

For the allocation problem in Eq.~\eqref{eq-allocation-problem}, let $\lambda_{\mathrm{norm}}\in\mathbb R$ enforce $\sum_\alpha a_\alpha=1$ and let $\nu\in\mathbb R^L$ contain the nonnegative multipliers for $a_\alpha\ge0$.
The Lagrangian is $\Phi(a)+\lambda_{\mathrm{norm}}(\sum_\alpha a_\alpha-1)-\nu^Ta$.
At an allocation with finite $\Phi(a)$, the KKT conditions are
\begin{equation}
\begin{aligned}
-\omega_\alpha(a)+\lambda_{\mathrm{norm}}-\nu_\alpha&=0,
&&\forall \alpha,\\
a_\alpha&\ge0,\ \nu_\alpha\ge0,\ a_\alpha\nu_\alpha=0,
&&\forall \alpha,\\
\sum_\alpha a_\alpha&=1.
\end{aligned}
\label{eq-allocation-kkt-system}
\end{equation}
The first row gives stationarity, and $a_\alpha\nu_\alpha=0$ expresses complementary slackness for the nonnegativity constraint.
Strictly positive shot fractions give a feasible allocation with $K(a)\succ0$, so these conditions are necessary for an optimum.
Convexity makes them sufficient.
Multiplying the stationarity conditions by $a_\alpha$ and summing gives $\lambda_{\mathrm{norm}}=\sum_\alpha a_\alpha \omega_\alpha(a)=\Phi(a)$, which yields Eq.~\eqref{eq-kkt}.

Let $\mathcal D$ be a finite collection of QWC groups not already in $\mathcal G$, with fixed positive-definite model covariance matrices.
The allocation conditions imply
\begin{equation}
\Phi_{\mathcal G\cup\mathcal D}^\star<\Phi_{\mathcal G}^\star
\ \Longleftrightarrow \
\exists G\in\mathcal D:\ \omega_G(a^\star)>\Phi_{\mathcal G}^\star.
\label{eq-admission}
\end{equation}
Extend the optimal allocation $a^\star\in\mathbb R^L$ by one zero entry for each group in $\mathcal D$, giving
\begin{equation}
\widetilde a=((a^\star)^T,0,\ldots,0)^T\in\mathbb R^{L+|\mathcal D|}.
\end{equation}
If any $G\in\mathcal D$ satisfies $\omega_G(a^\star)>\Phi_{\mathcal G}^\star$, Eq.~\eqref{eq-direction} gives a negative derivative when shots are transferred to that group.
A sufficiently small transfer keeps the corresponding matrix $K$ positive definite and decreases the objective.
If every additional group instead satisfies $\omega_G(a^\star)\le\Phi_{\mathcal G}^\star$, then $\widetilde a$ satisfies all KKT conditions for the combined group set.
Its objective is therefore already optimal even when all groups in $\mathcal D$ are available.

\subsection{Score increments and linear prediction}

For a group $G$ and an additional compatible term $i$, the covariance matrix $\Gamma_{G\cup\{i\}}\in\mathbb R^{(|G|+1)\times(|G|+1)}$ has the block form below.
Here $\Gamma_G\in\mathbb R^{|G|\times|G|}$, $u\in\mathbb R^{|G|}$, and $\Gamma_{ii}\in\mathbb R$ are the covariance quantities defined in Section~\ref{sec-construction}.
Define the linear prediction coefficients $b\in\mathbb R^{|G|}$ and residual variance $\sigma_{\mathrm{res}}^2\in\mathbb R$ by
\begin{equation}
\Gamma_{G\cup\{i\}}=
\begin{pmatrix}\Gamma_G&u\\u^T&\Gamma_{ii}\end{pmatrix},
\ b=\Gamma_G^{-1}u,
\ \sigma_{\mathrm{res}}^2=\Gamma_{ii}-u^T\Gamma_G^{-1}u>0.
\label{eq-nested}
\end{equation}
Positive definiteness of the enlarged covariance matrix gives $\sigma_{\mathrm{res}}^2>0$.
In a matrix inequality, $\preccurlyeq$ means that the matrix on the right minus the matrix on the left is positive semidefinite.
The block inverse identity gives
\begin{equation}
\Gamma_{G\cup\{i\}}^{-1}-
\begin{pmatrix}\Gamma_G^{-1}&0\\0&0\end{pmatrix}
=\frac{1}{\sigma_{\mathrm{res}}^2}
\begin{pmatrix}b\\-1\end{pmatrix}
\begin{pmatrix}b\\-1\end{pmatrix}^{T}
\succcurlyeq0.
\label{eq-rank-one}
\end{equation}
The off-diagonal zero blocks have sizes $|G|\times1$ and $1\times|G|$, and the final zero is a scalar.
The column vector $(b^T,-1)^T\in\mathbb R^{|G|+1}$ makes the right side a positive semidefinite matrix of size $(|G|+1)\times(|G|+1)$.
Evaluating this quadratic form at $(\mu_G^T,\mu_i)^T\in\mathbb R^{|G|+1}$ proves Eq.~\eqref{eq-increment}.
Placing the entries in the rows and columns indexed by the corresponding Pauli operators also gives $W_{G\cup\{i\}}-W_G\succcurlyeq0$, an inequality between $N_P\times N_P$ matrices.

Let $Z_G\in\mathbb R^{|G|}$ and $Z_i\in\mathbb R$ be a random vector and a random variable, respectively, both with zero mean. Their joint covariance matrix is given in Eq.~\eqref{eq-nested}.
The scalar residual $R_i=Z_i-b^TZ_G$ satisfies $\operatorname{Cov}(Z_G,R_i)=0\in\mathbb R^{|G|}$ and $\operatorname{Var}(R_i)=\sigma_{\mathrm{res}}^2$.
Thus $b^TZ_G$ is the best linear prediction of $Z_i$ under the specified covariance model, and $\sigma_{\mathrm{res}}^2$ is its residual variance.
This argument uses only second moments. Interpreting $\sigma_{\mathrm{res}}^2$ as a conditional variance requires additional assumptions.

\section{Shot allocation}\label{app-numerical}

This appendix gives the optimality check and convergence analysis for the Frank--Wolfe allocation iteration in Algorithm~\ref{alg-optimization}.
Section~\ref{sec-construction-cost} gives its arithmetic cost.

\subsection{Allocation update and convergence}\label{app-convergence}

Let $a^{(t)}\in\Delta_L\subset\mathbb R^L$ denote the allocation at iteration $t=0,1,\ldots$ for a fixed group set.
Denote the selected group index by $q_t\in\{1,\ldots,L\}$ and the transferred fraction by $\gamma_t\in[0,1]$.
The vector $e_{q_t}\in\mathbb R^L$ assigns all shots to the selected group.
Algorithm~\ref{alg-optimization} uses Eq.~\eqref{eq-assignment} and the Frank--Wolfe update~\cite{wynn1970sequential,jaggi2013revisiting}
\begin{equation}
q_t\in\operatorname{arg\,max}_{\alpha\in\{1,\ldots,L\}}\omega_\alpha(a^{(t)}),
\ a^{(t+1)}=(1-\gamma_t)a^{(t)}+\gamma_t e_{q_t}.
\label{eq-fw-update}
\end{equation}
Exact line minimization chooses $\gamma_t$ to minimize $\Phi$ along this transfer, assigning $\Phi=+\infty$ if its endpoint has a singular $K$.

For a prescribed dimensionless relative tolerance $\varepsilon_{\mathrm{rel}}>0$, define the score gap and the associated stopping condition as
\begin{equation}
g(a)=\max_\alpha \omega_\alpha(a)-\Phi(a),
\ \frac{g(a)}{\Phi(a)}\le\varepsilon_{\mathrm{rel}}.
\label{eq-inner-stop}
\end{equation}
Since the mean score weighted by the shot fractions is $\Phi(a)$, $g(a)$ is nonnegative and vanishes exactly when Eq.~\eqref{eq-kkt} holds.
The score gap can also assess allocations returned by other solvers. The convergence rate below, however, concerns the stated Frank--Wolfe iteration.
The proof below shows that $g(a)$ bounds the remaining objective error when the coefficient solve is exact.
Using this bound numerically requires control of the residual in $K(a)\mu=c$ and verification of the coefficient equalities and shot fraction normalization.

For the convergence analysis, assume that $K(a^{(0)})\succ0$ and that coefficient assignment and line minimization are exact.
Write $\Phi_0=\Phi(a^{(0)})$.
Let $\pi(a)=(\pi_1(a),\ldots,\pi_{N_P}(a))^T\in\mathbb R^{N_P}$, where $\pi_j(a)=\sum_{\alpha:j\in G_\alpha}a_\alpha$ is the fraction of shots that measure term $j$.
Define the positive covariance bounds $\ell=\min_\alpha\lambda_{\min}(\Gamma_\alpha^{-1})$ and $\beta=\max_\alpha\lambda_{\max}(\Gamma_\alpha^{-1})$, where $\lambda_{\max}$ denotes the largest eigenvalue.
Write $\operatorname{diag}(\pi(a))\in\mathbb R^{N_P\times N_P}$ for the diagonal matrix whose entries are $\pi_j(a)$.
Since each $S_\alpha$ selects the entries belonging to $G_\alpha$,
\begin{equation}
\ell\operatorname{diag}(\pi(a))\preccurlyeq K(a)
\preccurlyeq\beta\operatorname{diag}(\pi(a)).
\label{eq-spectral-sandwich}
\end{equation}
Consequently, $\Phi(a)\ge\sum_j c_j^2/(\beta \pi_j(a))$.
For every allocation satisfying $\Phi(a)\le\Phi_0$, the fraction measuring term $j$ satisfies $\pi_j(a)\ge c_j^2/(\beta\Phi_0)$, and hence
\begin{equation}
K(a)\succcurlyeq\kappa I_{N_P},\
\kappa=\frac{\ell\min_j c_j^2}{\beta\Phi_0}>0.
\label{eq-spectral-floor}
\end{equation}
Exact line minimization does not increase $\Phi$, so every allocation generated from $a^{(0)}$ satisfies this bound.

For the selected group $q$ with the highest score, define $K_\gamma=K(a)+\gamma(W_q-K(a))\in\mathbb R^{N_P\times N_P}$ for $0\le\gamma\le1/2$.
This is the matrix corresponding to the allocation $(1-\gamma)a+\gamma e_q$.
For matrices, $\|\cdot\|_2$ denotes the largest singular value.
The bounds above give $K_\gamma\succcurlyeq\kappa I_{N_P}/2$ and $\|W_q-K(a)\|_2\le2\beta$.
Differentiating $\Phi((1-\gamma)a+\gamma e_q)$ twice with respect to $\gamma$ bounds its second derivative from above by the positive scalar $C_{\mathrm{FW}}=64\beta^2\|c\|_2^2/\kappa^3$.
Therefore
\begin{equation}
\Phi((1-\gamma)a+\gamma e_q)
\le\Phi(a)-\gamma g(a)+\frac{C_{\mathrm{FW}}}{2}\gamma^2,
\ 0\le\gamma\le\frac12.
\label{eq-fw-comparison}
\end{equation}
Exact line minimization yields an objective value no greater than that of any comparison step, including when the full segment has a singular endpoint.

\begin{theorem}[Convergence for a fixed group set]\label{thm-fw}
Under the assumptions above, let $h_t=\Phi(a^{(t)})-\Phi_{\mathcal G}^\star$ be the objective error and set $g_t=g(a^{(t)})$ and $A=\max\{3\Phi_0,2C_{\mathrm{FW}}\}$.
Then, for every iteration $t\ge0$ and every integer $n\ge2$,
\begin{equation}
h_t\le\frac{A}{t+3},
\ \min_{n\le t<2n}g_t\le\frac{A}{n+3}+\frac{C_{\mathrm{FW}}}{2n}.
\label{eq-fw-rate}
\end{equation}
\end{theorem}

\begin{proof}
Let $a^\star\in\Delta_L$ be an optimal allocation.
The gradient $\nabla\Phi(a)\in\mathbb R^L$ collects the allocation derivatives in Eq.~\eqref{eq-gradient}, with component $-\omega_\alpha(a)$ in position $\alpha$.
Convexity gives
$0\le h_t\le-\nabla\Phi(a^{(t)})^T(a^\star-a^{(t)})\le g_t$, because the maximum group score is at least its average under $a^\star$.
Using this inequality and $\gamma=2/(t+4)$ in Eq.~\eqref{eq-fw-comparison} gives
\[
h_{t+1}\le\frac{t+2}{t+4}h_t+\frac{2C_{\mathrm{FW}}}{(t+4)^2}.
\]
The first bound follows by induction from $h_0\le A/3$ and $A\ge2C_{\mathrm{FW}}$.
For the second bound, use $\gamma=1/n$ at each iteration $n\le t<2n$ and sum the resulting decreases.
The average of $g_t$ over this interval is at most $h_n+C_{\mathrm{FW}}/(2n)$. The same bound therefore applies to the minimum value of $g_t$ over the interval.
Substitution of the first bound gives Eq.~\eqref{eq-fw-rate}.
\end{proof}

For a prescribed absolute objective tolerance $\varepsilon_{\mathrm{abs}}>0$, the sequence therefore reaches an iterate with $g_t\le\varepsilon_{\mathrm{abs}}$ in $O((A+C_{\mathrm{FW}})/\varepsilon_{\mathrm{abs}})$ iterations.
This tolerance has the same units as $\Phi$.

\section{Experimental details}\label{app-protocol}

\subsection{Molecular systems and reference states}\label{app-protocol-inputs}

All five calculations used the complete STO-3G spin-orbital spaces, with no frozen core or qubit tapering, and the Bravyi--Kitaev mapping. 
The nuclear geometries are specified in Section~\ref{sec-molecular-systems}.
Identity energy offsets contribute no sampling variance and were omitted from the variance calculations.
The reference state was obtained by diagonalizing the molecular Hamiltonian in the sector with $N_e$ electrons and $N_\alpha=N_\beta$, where $N_\alpha$ and $N_\beta$ are the numbers of electrons with alpha and beta spin, respectively.

The CISD calculation used a reference Slater determinant, a many-electron state defined as the antisymmetrized product of its occupied spin orbitals. 
The first $N_e$ spin orbitals in the interleaved alpha/beta ordering were occupied, and all remaining spin orbitals were empty.
Single and double excitations generated further determinants by replacing one or two occupied spin orbitals with unoccupied ones. 
The CISD state was the eigenvector with the lowest energy after restricting the Hamiltonian to the subspace spanned by the reference determinant and these excited determinants within the same electron-number sector.
Covariances were evaluated from the corresponding state vectors. 
Group construction and optimization used one of these fixed covariance sources, while reported physical variances used the reference state.

\subsection{Covariance model and numerical optimization}\label{app-protocol-numerics}

For joint optimization of coefficients and shot fractions, including the ICS and SOCP implementations and VGMC, the design blocks were
\begin{equation}
\Gamma_G=C_G+\lambda_{\mathrm{cov}} I_{|G|},\ \lambda_{\mathrm{cov}}=10^{-8},
\label{eq-design-regularization}
\end{equation}
where $C_G\in\mathbb R^{|G|\times|G|}$ is the covariance block of the exact or CISD state and $\lambda_{\mathrm{cov}}>0$ is the ridge parameter.
The same ridge was added to every diagonal. For nested groups, the covariance matrix of the smaller group therefore remained a principal submatrix of that of the larger group.
Thus \textit{exact-covariance design} identifies the source of $C_G$ before regularization.

The reduced solvers used L-BFGS-B, with coefficient recovery through Eq.~\eqref{eq-assignment}.
They parameterized positive numerical allocations as
\begin{equation}
a_\alpha(\theta)=\frac{\theta_\alpha^2+\delta}{\sum_\beta(\theta_\beta^2+\delta)},
\ \theta\in\mathbb R^L,\ \delta=10^{-15}.
\label{eq-numerical-allocation}
\end{equation}
Here $\delta$ is a floor on unnormalized weights. The theoretical problem remains defined on the closed simplex.

Numerical optimality was assessed in the original allocation coordinates.
For feasible returned coefficients, let $U=\Psi(x,a)>0$ be the scalar primal objective. A dual objective is obtained by scaling the multiplier vector $\mu\in\mathbb R^{N_P}$ so that its largest group score satisfies the conic dual constraint in Appendix~\ref{app-socp-duality}.
In exact arithmetic, for a finite set $\mathcal P$ containing the current groups, this gives
\begin{equation}
\Phi_{\mathcal P}^{\mathrm{lb}}=\frac{(c^T\mu)^2}{\max_{G\in\mathcal P}\omega_G},
\ \varepsilon_{\mathrm{pd}}=\frac{U}{\Phi_{\mathcal P}^{\mathrm{lb}}}-1.
\label{eq-numerical-primal-dual-gap}
\end{equation}
Here $\Phi_{\mathcal P}^{\mathrm{lb}}$ is a scalar lower bound on $\Phi_{\mathcal P}^\star$ and $\varepsilon_{\mathrm{pd}}$ is a dimensionless relative primal--dual gap.
For $\mathcal P=\mathcal G$, the numerical implementation compares this gap with the prescribed tolerance $\varepsilon_{\mathrm{rel}}$ in Algorithm~\ref{alg-optimization}.
For a feasible estimator and a positive dual lower bound, the gap test implies $\Phi_{\mathcal G}^\star\le \Psi(x,a)\le(1+\varepsilon_{\mathrm{rel}})\Phi_{\mathcal G}^\star$ in exact arithmetic.
Coefficient conservation was enforced before evaluating these quantities, and numerical safety margins were applied to account for residuals and roundoff.
The resulting gap estimates are numerical diagnostics computed in floating point.
With exact coefficient recovery on the current group set and without safety padding, Eq.~\eqref{eq-numerical-primal-dual-gap} equals the relative score gap $g(a)/\Phi(a)$ in Appendix~\ref{app-convergence}.
For the current group set, the final target was $\varepsilon_{\mathrm{rel}}=10^{-4}$, with a tolerance of $10^{-9}$ for coefficient conservation.
\textcolor{black}{
Expansion solves first targeted $10^{-3}$ and were refined to $10^{-4}$ at the evaluated addition counts. Unresolved candidate boundaries allowed further allocation refinement to $10^{-5}$ and $10^{-6}$.}
ICS was run for five updates for the comparison in Table~\ref{tab-molecular-variance}~\cite{Kottmann_Tequila_A_platform_2020}.
The SOCP implementation used CLARABEL with absolute and relative gap and feasibility tolerances $10^{-9}$ and at most 500 iterations.
\bibliographystyle{apsrev4-2}
\bibliography{references}

@dataset{vgmc_2026_v1,
  author       = {Zhao, Taotao and Yao, Qianjun and Sun, Mengdi and Li, He},
  title        = {From optimization to construction: Expanding quantum measurement designs},
  year         = 2026,
  publisher    = {Zenodo},
  doi          = {10.5281/zenodo.24068546},
  url          = {https://doi.org/10.5281/zenodo.24068546},
}

@book{boyd2004convex,
  author = {Boyd, Stephen and Vandenberghe, Lieven},
  title = {Convex Optimization},
  publisher = {Cambridge University Press},
  year = {2004},
  url = {https://web.stanford.edu/~boyd/cvxbook/}
}

@article{byrd1995limited,
  title={A limited memory algorithm for bound constrained optimization},
  author={Byrd, Richard H and Lu, Peihuang and Nocedal, Jorge and Zhu, Ciyou},
  journal={SIAM Journal on scientific computing},
  volume={16},
  number={5},
  pages={1190--1208},
  year={1995},
  publisher={SIAM},
  eprint = {}
}

@article{lobo1998applications,
  author = {Lobo, Miguel Sousa and Vandenberghe, Lieven and Boyd, Stephen and Lebret, Herv{\'e}},
  title = {Applications of Second-Order Cone Programming},
  journal = {Linear Algebra and its Applications},
  volume = {284},
  pages = {193--228},
  year = {1998},
  url = {https://stanford.edu/~boyd/papers/socp.html}
}

@article{bravyi2002fermionic,
  author  = {Bravyi, Sergey B. and Kitaev, Alexei Yu.},
  title   = {{Fermionic Quantum Computation}},
  journal = {Annals of Physics},
  volume  = {298},
  number  = {1},
  pages   = {210--226},
  year    = {2002},
  doi     = {10.1006/aphy.2002.6254},
}

@article{seeley2012bravyi,
  author  = {Seeley, Jacob T. and Richard, Martin J. and Love, Peter J.},
  title   = {{The Bravyi--Kitaev transformation for quantum computation of electronic structure}},
  journal = {The Journal of Chemical Physics},
  volume  = {137},
  number  = {22},
  pages   = {224109},
  year    = {2012},
  doi     = {10.1063/1.4768229},
}

@article{crawford2021efficient,
  author  = {Crawford, Ophelia and van Straaten, Barnaby and Wang, Daochen and Parks, Thomas and Campbell, Earl and Brierley, Stephen},
  title   = {Efficient Quantum Measurement of {P}auli Operators in the Presence of Finite Sampling Error},
  journal = {Quantum},
  volume  = {5},
  pages   = {385},
  year    = {2021},
  doi     = {10.22331/q-2021-01-20-385},
}

@article{huidobro2026initializations,
  author  = {Huidobro-Meezs, Isaac L. and Vargas-Hern{\'a}ndez, Rodrigo A.},
  title   = {Reducing Quantum Measurements in Qubit-Based Overlapping Grouping Methods for Quantum Energy Estimation through Better Initializations},
  journal = {arXiv preprint arXiv:2607.02794v2},
  note    = {Version 2, 9 September 2026},
  url     = {https://arxiv.org/abs/2607.02794v2},
  year    = {2026},
  doi     = {10.48550/arXiv.2607.02794},
}

@inproceedings{jaggi2013revisiting,
  url = {https://proceedings.mlr.press/v28/jaggi13.html},
  author    = {Jaggi, Martin},
  title     = {Revisiting Frank--Wolfe: Projection-Free Sparse Convex Optimization},
  booktitle = {Proceedings of the 30th International Conference on Machine Learning},
  series    = {Proceedings of Machine Learning Research},
  volume    = {28},
  pages     = {427--435},
  year      = {2013},
}

@article{mcclean2016theory,
  author  = {McClean, Jarrod R. and Romero, Jonathan and Babbush, Ryan and Aspuru-Guzik, Al{\'a}n},
  title   = {The Theory of Variational Hybrid Quantum-Classical Algorithms},
  journal = {New Journal of Physics},
  volume  = {18},
  pages   = {023023},
  year    = {2016},
  doi     = {10.1088/1367-2630/18/2/023023},
}

@article{peruzzo2014vqe,
  author  = {Peruzzo, Alberto and McClean, Jarrod and Shadbolt, Peter and Yung, Man-Hong and Zhou, Xiao-Qi and Love, Peter J. and Aspuru-Guzik, Al{\'a}n and O'Brien, Jeremy L.},
  title   = {A Variational Eigenvalue Solver on a Photonic Quantum Processor},
  journal = {Nature Communications},
  volume  = {5},
  pages   = {4213},
  year    = {2014},
  doi     = {10.1038/ncomms5213},
}

@article{rowland2026overlapped,
  author  = {Rowland, Jeremiah and Sarkar, Rahul and Sawaya, Nicolas P. D. and Tubman, Norm M. and LaRose, Ryan},
  title   = {Overlapped Groupings for Quantum Energy Estimation: Maximal Variance Reduction and Deterministic Algorithms for Reducing Variance},
  journal = {arXiv preprint arXiv:2604.07156},
  year    = {2026},
  doi     = {10.48550/arXiv.2604.07156},
}

@article{sagnol2011multiresponse,
  author  = {Sagnol, Guillaume},
  title   = {Computing Optimal Designs of Multiresponse Experiments Reduces to Second-Order Cone Programming},
  journal = {Journal of Statistical Planning and Inference},
  volume  = {141},
  number  = {5},
  pages   = {1684--1708},
  year    = {2011},
  doi     = {10.1016/j.jspi.2010.11.031},
}

@article{verteletskyi2020minimumclique,
  author  = {Verteletskyi, Vladyslav and Yen, Tzu-Ching and Izmaylov, Artur F.},
  title   = {Measurement Optimization in the Variational Quantum Eigensolver Using a Minimum Clique Cover},
  journal = {Journal of Chemical Physics},
  volume  = {152},
  pages   = {124114},
  year    = {2020},
  doi     = {10.1063/1.5141458},
}

@article{wu2023ogm,
  author  = {Wu, Bujiao and Sun, Jinzhao and Huang, Qi and Yuan, Xiao},
  title   = {Overlapped Grouping Measurement: A Unified Framework for Measuring Quantum States},
  journal = {Quantum},
  volume  = {7},
  pages   = {896},
  year    = {2023},
  doi     = {10.22331/q-2023-01-13-896},
}

@article{wynn1970sequential,
  author  = {Wynn, Henry P.},
  title   = {The Sequential Generation of {$D$}-Optimum Experimental Designs},
  journal = {The Annals of Mathematical Statistics},
  volume  = {41},
  number  = {5},
  pages   = {1655--1664},
  year    = {1970},
  doi     = {10.1214/aoms/1177696809},
}

@article{yen2023deterministic,
  author  = {Yen, Tzu-Ching and Ganeshram, Aadithya and Izmaylov, Artur F.},
  title   = {Deterministic Improvements of Quantum Measurements with Grouping of Compatible Operators, Non-Local Transformations, and Covariance Estimates},
  journal = {npj Quantum Information},
  volume  = {9},
  pages   = {14},
  year    = {2023},
  doi     = {10.1038/s41534-023-00683-y},
}

@article{wecker2015solving,
  author  = {Wecker, Dave and Hastings, Matthew B. and Troyer, Matthias},
  title   = {{Progress towards practical quantum variational algorithms}},
  journal = {Physical Review A},
  volume  = {92},
  number  = {4},
  pages   = {042303},
  year    = {2015},
  doi     = {10.1103/PhysRevA.92.042303},
}

@article{yen2020measuring,
  author  = {Yen, Tzu-Ching and Verteletskyi, Vladyslav and Izmaylov, Artur F.},
  title   = {{Measuring All Compatible Operators in One Series of Single-Qubit Measurements Using Unitary Transformations}},
  journal = {Journal of Chemical Theory and Computation},
  volume  = {16},
  number  = {4},
  pages   = {2400--2409},
  year    = {2020},
  doi     = {10.1021/acs.jctc.0c00008},
}

@article{huang2020classical,
  author  = {Huang, Hsin-Yuan and Kueng, Richard and Preskill, John},
  title   = {{Predicting many properties of a quantum system from very few measurements}},
  journal = {Nature Physics},
  volume  = {16},
  number  = {10},
  pages   = {1050--1057},
  year    = {2020},
  doi     = {10.1038/s41567-020-0932-7},
}

@article{huang2021derandomization,
  author  = {Huang, Hsin-Yuan and Kueng, Richard and Preskill, John},
  title   = {{Efficient Estimation of Pauli Observables by Derandomization}},
  journal = {Physical Review Letters},
  volume  = {127},
  number  = {3},
  pages   = {030503},
  year    = {2021},
  doi     = {10.1103/PhysRevLett.127.030503},
}

@article{hadfield2022biased,
  author  = {Hadfield, Charles and Bravyi, Sergey and Raymond, Rudy and Mezzacapo, Antonio},
  title   = {{Measurements of Quantum Hamiltonians with Locally-Biased Classical Shadows}},
  journal = {Communications in Mathematical Physics},
  volume  = {391},
  number  = {3},
  pages   = {951--967},
  year    = {2022},
  doi     = {10.1007/s00220-022-04343-8},
}

@article{cao2019chemistry,
  author  = {Cao, Yudong and Romero, Jonathan and Olson, Jonathan P. and Degroote, Matthias and Johnson, Peter D. and Kieferová, Mária and Kivlichan, Ian D. and Menke, Tim and Peropadre, Borja and Sawaya, Nicolas P. D. and Sim, Sukin and Veis, Libor and Aspuru-Guzik, Alán},
  title   = {{Quantum Chemistry in the Age of Quantum Computing}},
  journal = {Chemical Reviews},
  volume  = {119},
  number  = {19},
  pages   = {10856--10915},
  year    = {2019},
  doi     = {10.1021/acs.chemrev.8b00803},
}

@article{mcardle2020chemistry,
  author  = {McArdle, Sam and Endo, Suguru and Aspuru-Guzik, Alán and Benjamin, Simon C. and Yuan, Xiao},
  title   = {{Quantum computational chemistry}},
  journal = {Reviews of Modern Physics},
  volume  = {92},
  number  = {1},
  pages   = {015003},
  year    = {2020},
  doi     = {10.1103/RevModPhys.92.015003},
}

@article{cerezo2021variational,
  author  = {Cerezo, M. and Arrasmith, Andrew and Babbush, Ryan and Benjamin, Simon C. and Endo, Suguru and Fujii, Keisuke and McClean, Jarrod R. and Mitarai, Kosuke and Yuan, Xiao and Cincio, Lukasz and Coles, Patrick J.},
  title   = {{Variational quantum algorithms}},
  journal = {Nature Reviews Physics},
  volume  = {3},
  number  = {9},
  pages   = {625--644},
  year    = {2021},
  doi     = {10.1038/s42254-021-00348-9},
}

@article{bharti2022noisy,
  author  = {Bharti, Kishor and Cervera-Lierta, Alba and Kyaw, Thi Ha and Haug, Tobias and Alperin-Lea, Sumner and Anand, Abhinav and Degroote, Matthias and Heimonen, Hermanni and Kottmann, Jakob S. and Menke, Tim and Mok, Wai-Keong and Sim, Sukin and Kwek, Leong-Chuan and Aspuru-Guzik, Alán},
  title   = {{Noisy intermediate-scale quantum algorithms}},
  journal = {Reviews of Modern Physics},
  volume  = {94},
  number  = {1},
  pages   = {015004},
  year    = {2022},
  doi     = {10.1103/RevModPhys.94.015004},
}

@article{tilly2022vqe,
  author  = {Tilly, Jules and Chen, Hongxiang and Cao, Shuxiang and Picozzi, Dario and Setia, Kanav and Li, Ying and Grant, Edward and Wossnig, Leonard and Rungger, Ivan and Booth, George H. and Tennyson, Jonathan},
  title   = {{The Variational Quantum Eigensolver: A review of methods and best practices}},
  journal = {Physics Reports},
  volume  = {986},
  pages   = {1--128},
  year    = {2022},
  doi     = {10.1016/j.physrep.2022.08.003},
}

@article{elben2023randomized,
  author  = {Elben, Andreas and Flammia, Steven T. and Huang, Hsin-Yuan and Kueng, Richard and Preskill, John and Vermersch, Benoît and Zoller, Peter},
  title   = {{The randomized measurement toolbox}},
  journal = {Nature Reviews Physics},
  volume  = {5},
  number  = {1},
  pages   = {9--24},
  year    = {2023},
  doi     = {10.1038/s42254-022-00535-2},
}

@article{gonthier2022roadblock,
  author  = {Gonthier, Jérôme F. and Radin, Maxwell D. and Buda, Corneliu and Doskocil, Eric J. and Abuan, Clena M. and Romero, Jhonathan},
  title   = {{Measurements as a roadblock to near-term practical quantum advantage in chemistry: Resource analysis}},
  journal = {Physical Review Research},
  volume  = {4},
  number  = {3},
  pages   = {033154},
  year    = {2022},
  doi     = {10.1103/PhysRevResearch.4.033154},
}

@article{hamamura2020efficient,
  author  = {Hamamura, Ikko and Imamichi, Takashi},
  title   = {{Efficient evaluation of quantum observables using entangled measurements}},
  journal = {npj Quantum Information},
  volume  = {6},
  number  = {1},
  pages   = {56},
  year    = {2020},
  doi     = {10.1038/s41534-020-0284-2},
}

@article{burns2025galic,
  author  = {Burns, Matthew X. and Liu, Chenxu and Stein, Samuel and Peng, Bo and Kowalski, Karol and Li, Ang},
  title   = {{GALIC: Hybrid multi-qubitwise Pauli grouping for quantum computing measurement}},
  journal = {Quantum Science and Technology},
  volume  = {10},
  number  = {1},
  pages   = {015054},
  year    = {2025},
  doi     = {10.1088/2058-9565/ad9d74},
}

@article{dalfavero2025commutativity,
  author  = {DalFavero, Ben and Sarkar, Rahul and Rowland, Jeremiah and Camps, Daan and Sawaya, Nicolas P. D. and LaRose, Ryan},
  title   = {{Measurement reduction for expectation values via fine-grained commutativity}},
  journal = {Physical Review A},
  volume  = {112},
  number  = {5},
  pages   = {052407},
  year    = {2025},
  doi     = {10.1103/y6hl-gzr4},
}

@article{miller2024hardware,
  author  = {Miller, Daniel and Fischer, Laurin E. and Levi, Kyano and Kuehnke, Eric J. and Sokolov, Igor O. and Barkoutsos, Panagiotis Kl. and Eisert, Jens and Tavernelli, Ivano},
  title   = {{Hardware-tailored diagonalization circuits}},
  journal = {npj Quantum Information},
  volume  = {10},
  number  = {1},
  pages   = {122},
  year    = {2024},
  doi     = {10.1038/s41534-024-00901-1},
}

@article{Kottmann_Tequila_A_platform_2020,
  title={Tequila: A platform for rapid development of quantum algorithms},
  author={Kottmann, Jakob S and Alperin-Lea, Sumner and Tamayo-Mendoza, Teresa and Cervera-Lierta, Alba and Lavigne, Cyrille and Yen, Tzu-Ching and Verteletskyi, Vladyslav and Schleich, Philipp and Anand, Abhinav and Degroote, Matthias and others},
  journal={Quantum Science \& Technology},
  volume={6},
  number={2},
  pages={024009},
  year={2021},
  publisher={IOP Publishing},
  eprint={}
}
\end{document}